\documentclass[traditabstract]{aa} 
\usepackage{graphicx}
\usepackage{txfonts}
\begin{document}
   \title{Physical Properties of Asteroid (308635) 2005~YU$_{55}$ derived from
          multi-instrument infrared observations during a very close Earth-Approach}


   \author{
          T. G. M\"{u}ller \inst{1},
          T. Miyata \inst{2},
          C. Kiss \inst{3}
          M. A. Gurwell \inst{4}
          S. Hasegawa \inst{5},
          E. Vilenius \inst{1},
          S. Sako \inst{2},
          T. Kamizuka \inst{2},
          T. Nakamura \inst{6},
          K. Asano \inst{2},
          M. Uchiyama \inst{2},
          M. Konishi \inst{2},
          M. Yoneda \inst{7},
          T. Ootsubo \inst{8},
          F. Usui \inst{5},
          Y. Yoshii \inst{2},
          M. Kidger \inst{9},
          B. Altieri \inst{9},
          R. Lorente \inst{9},
          A. P\'al \inst{3},
          L. O'Rourke \inst{9},
          \and
          L. Metcalfe \inst{9}
          }

   \institute{   {Max-Planck-Institut f\"{u}r extraterrestrische Physik, Giessenbachstra{\ss}e,
                 Postfach 1312, 85741 Garching, Germany
                 }
          \and
                 {Institute of Astronomy, School of Science, the University of Tokyo,
                 2-21-1 Osawa, Mitaka, Tokyo 181-0015, Japan
                 \email{tmiyata@ioa.s.u-tokyo.ac.jp}
                 }
          \and
                 {Konkoly Observatory, Research Center for Astronomy and
                 Earth Sciences, Hungarian Academy of Sciences;
                 Konkoly Thege 15-17, H-1121 Budapest, Hungary
                 }
          \and
                 {Harvard-Smithsonian Center for Astrophysics,
                 60 Garden Street, Cambridge, MA 02138, USA
                 }
          \and
                 {Institute of Space and Astronautical Science, Japan Aerospace Exploration Agency,
                 3-1-1 Yoshinodai, Sagamihara, Kanagawa 229-8510, Japan
                 }
          \and
                 {Department of Astronomy, Graduate School of Science, The University of Tokyo,
                 Hongo 7-3-1, Bunkyo-ku, Tokyo 113-0033, Japan
                 }
          \and
                 {Planetary Plasma and Atmospheric Research Center, Tohoku University,
                 Aramaki, Aoba-ku, Sendai 980-8578, Japan
                 }
          \and
                 {Astronomical Institute, Graduate School of Science, Tohoku University,
                 Aramaki, Aoba-ku, Sendai 980-8578, Japan
                 }
          \and
                 {European Space Astronomy Centre (ESAC), European Space Agency,
                 Apartado de Correos 78, 28691 Villanueva de la Ca\~nada,
                 Madrid, Spain
                 }
          }

   \date{Received ; accepted }

\abstract{The near-Earth asteroid (308635) 2005~YU$_{55}$ is a potentially hazardous asteroid
          which was discovered in 2005 and passed Earth on November 8$^{th}$ 2011 at 0.85 lunar
	  distances. This was the closest known approach by an asteroid of several hundred
	  metre diameter since 1976 when a similar size object passed at 0.5 lunar distances.
	  We observed 2005~YU$_{55}$ from ground with a recently developed mid-IR camera
	  (miniTAO/MAX38) in N- and Q-band and with the Submillimeter Array (SMA) at 1.3\,mm.
	  In addition, we obtained space observations with Herschel\thanks{{\it Herschel} is an
          ESA space observatory with science instruments provided by European-led Principal
          Investigator consortia and with important participation from NASA.}/PACS at 70, 100, and 160\,$\mu$m.
          Our thermal measurements cover a wide range of wavelengths from 8.9\,$\mu$m to 1.3\,mm and were
          taken after opposition at phase angles between -97$^{\circ}$ and -18$^{\circ}$.
	  We performed a radiometric analysis via a thermophysical model and combined our derived properties
	  with results from radar, adaptive optics, lightcurve observations, speckle and auxiliary thermal data.
	  We find that (308635)~2005~YU$_{55}$ has an almost spherical shape with an effective diameter of
	  300 to 312\,m and a geometric albedo p$_V$ of 0.055 to 0.075. Its spin-axis is oriented
	  towards celestial directions ($\lambda_{ecl}$, $\beta_{ecl}$) = (60$^{\circ}$ $\pm$ 30$^{\circ}$,
	  -60$^{\circ}$ $\pm$ 15$^{\circ}$), which means it has a retrograde sense of rotation.
	  The analysis of all available data combined revealed a discrepancy with the radar-derived size.
	  Our radiometric analysis of the thermal data together with the problem to find a unique rotation
	  period might be connected to a non-principal axis rotation. A low to intermediate level of surface roughness (r.m.s.\ of surface slopes
	  in the range 0.1 - 0.3) is required to explain the available thermal measurements.
	  We found a thermal inertia in the range 350-800\,Jm$^{-2}$s$^{-0.5}$K$^{-1}$,
	  very similar to the rubble-pile asteroid (25143)~Itokawa and indicating a mixture of low conductivity
	  fine regolith with larger rocks and boulders of high thermal inertia on the surface. 
	  }


\authorrunning{M\"uller et al.}
\titlerunning{MIR Observations of (308635) 2005~YU$_{55}$}

   \maketitle
%

\section{Introduction}

The Apollo- and C-type asteroid (308635) 2005~YU$_{55}$ is on a
Mars-Earth-Venus crossing orbit\footnote{2005, M.P.E.C. 2005-Y47:\\
{\tt http:\-//www.minorplanetcenter.org/\-mpec/\-K05/\-K05Y47.html}}
(Vodniza \& Pereira \cite{vodniza10}; Hicks et al.\ \cite{hicks10};
Somers et al.\ \cite{somers10}).
Arecibo radar measurements in April 2010
have shown that 2005~YU$_{55}$ is a very dark, nearly spherical object\footnote{NASA Near Earth Object Program News:\\
{\tt http:\-//neo.jpl.nasa.gov/\-news/\-news171.html}}.
They estimated a diameter of about 400\,m, in contradiction to earlier
calculations based on the V-magnitude in combination with a low
albedo which led to a diameter of only 250\,m.

2005~YU$_{55}$ had a very close Earth approach
in November 2011 when it passed within 0.85 lunar distances (0.85\,LD)
of the Earth. Later, in January 2029, the asteroid will pass about 0.0023\,AU (equivalent
to 0.89\,LD) from Venus. This close encounter with Venus will determine
how close the object will pass the Earth in 2041 and 2045\footnote{http://echo.jpl.nasa.gov/asteroids/2005YU55/2005YU55\_planning.html}.
The JPL Horizons system gives the absolute
magnitude of 2005~YU$_{55}$ as H=21.1\,mag\footnote{JPL Horizons:
{\tt http:\-//ssd.jpl.nasa.gov/\-horizons.cgi}}.
No other asteroid with H$<$23\,mag has been observed before to pass
inside 1\,LD. According to recent orbit simulations it does not pose any
risk of an impact with Earth for the next 100 years\footnote{JPL's NEO
Radar Detection Program Webpage:\\ {\tt http:\-//echo.jpl.nasa.gov/\-asteroids/index.html}}.
The closest recorded approach by an asteroid of similar characteristics was
that of 2004~XP$_{14}$ (H=19.4\,mag) to 1.1\,LD on 2006 July 3, hence
the encounter with 2005~YU$_{55}$ was an exceptional event.

The close Earth approach in November 2011 offered a several day observing
opportunity from ground and also a brief ($\sim$16\,h) observing
window for the Herschel Space Observatory located in the Lagrangian point L2
at about 1.5 Mio km from Earth. This was a unique opportunity to study
a potentially hazardous asteroid (PHA) in great detail to derive physical
and thermal properties which are needed to make long-term orbit predictions
and to improve our knowledge on Apollo asteroids in general.
We observed this near-Earth asteroid from ground at mid-Infrared N- and Q-band
(miniTAO/MAX38 camera), at millimetre wavelength (Smithsonian Astrophysical Observatory
Submillimeter Array, or SMA) and from space with Herschel-PACS at far-infrared
wavelengths. We present our observations (Section \ref{sec:obs}), the thermophysical
model (TPM) analysis (Section \ref{sec:tpm}) and discuss the results (Section
\ref{sec:dis}). In this work, we also considered a set of auxiliary data
(radar, optical, UV and thermal measurements) which were only available via
unrefereed abstracts, astronomical circulars and telegrams.

\section{Observations}
\label{sec:obs}

\subsection{Groundbased mid-IR observations with MAX38}

\begin{figure}[h!tb]
 \rotatebox{0}{\resizebox{\hsize}{!}{\includegraphics{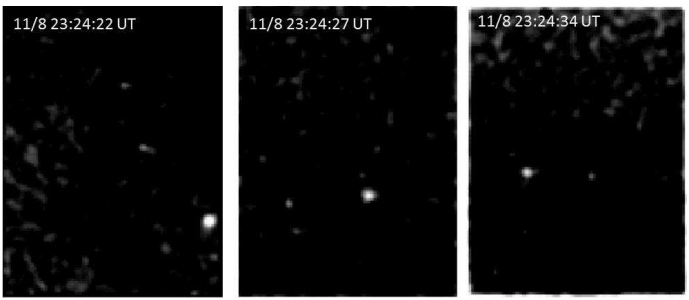}}}
  \caption{Mid-infrared images of 2005~YU$_{55}$ obtained by the miniTAO/MAX38
           camera in the 18.7\,$\mu$m filter during the time of the closest
           Earth approach. North is up and west is right. The asteroid moved
           from right (west) to left (east).
     \label{fig:max38}}
\end{figure}

We observed the asteroid 2005~YU$_{55}$ in the period Nov.\ 8 - 10, 2011
with the mid-infrared camera
MAX38 (Miyata et al.\ \cite{miyata08}; Nakamura et al.\ \cite{nakamura10};
Asano et al.\ \cite{asano12}) attached on the mini-TAO 1~meter telescope
(Sako et al.\ \cite{sako08}) which is located at 5640\,m altitude on the
summit of Co.\ Chajnantor in Chile, which is part of the University of Tokyo
Atacama Observatory Project (PI: Yuzuru Yoshii; Yoshii et al.\ \cite{yoshii10}).
MAX38 has a 128$\times$128 Si:Sb BIB detector with a pixel scale of 1.26\,arcsec
and a field of view of 2$\times$2.5\,arcmin determined by the rectangular
field stop in the cold optics (the remaining 0.5$\times$2.5\,arcmin of the
detector array is used for spectroscopy).
The MAX38 observing periods (2011-Nov-08 23:04 to Nov-09 01:51 UT and
from Nov-09 23:56 to Nov-10 02:04 UT) covered the time of the closest approach
(2011-Nov-08 23:24 UT)
and about 24 hours later. The weather conditions were excellent through the observations.
Imaging observations in the 8.9\,$\mu$m ($\Delta\lambda$ $\sim$ 0.9\,$\mu$m),
12.2\,$\mu$m (0.5\,$\mu$m), and 18.7\,$\mu$m (0.9\,$\mu$m) bands were carried out.
$\alpha$~Tuc (IRAS22150-6030, HR~8502, HD~211416) was also observed after the observations of the
asteroid as a flux standard.
The absolute flux value of the standard star was obtained via the $\alpha$~Tuc
template spectrum (Cohen et al.\ \cite{cohen99}).

Since the distance to the asteroid from the Earth was very short, the asteroid had a
very high apparent motion on the sky. We pointed the telescope at repeated intervals
to follow the asteroid's movement. The intervals were set to 1 minute and 3 minutes
on Nov.\ 8 and 9, respectively. Normal sidereal tracking was applied in the period
between the telescope pointings. Images were taken at a frame rate of 3.8\,Hz with
an effective integration time of 0.197\,sec.
The frame rate is fast enough not to extend the image of the asteroid on each frame.
Chopping technique\footnote{the telescope's secondary mirror is oscillated between
two positions on the sky at a frequency of a few Hz.} was not applied because
background can be canceled out
with using frames just before or after an object frame. The observation parameters
are summarized in Table~\ref{tbl:obsparam} and three examples of reduced images
taken at the time of the closest approach are shown in Fig.~\ref{fig:max38}.

%
 \begin{center}
   \begin{table}[h!tb]
      \caption{MAX38 Observation Parameters. The first period includes
               the closest Earth approach.}
      \label{tbl:obsparam}
      \begin{tabular}{ccrrcc}
            \noalign{\smallskip}
            \hline
            \hline
            \noalign{\smallskip}
            \multicolumn{2}{c}{2011-Nov-.. (UT)}  &  Filter & \# of  & \multicolumn{2}{c}{airmass-range} \\
            Day Start  & End &  Band   & frames & 2005~YU$_{55}$ & $\alpha$~Tuc \\
            \noalign{\smallskip}
            \hline
            \noalign{\smallskip}
            08 23:04 & 00:15 & 18.7 & 20196 & 1.31-1.47 & 1.41-1.48 \\
            09 00:30 & 00:41 &  8.9 &  2550 & 1.51-1.55 & 1.29 \\
            09 01:11 & 01:24 & 12.2 &  1734 & 1.65-1.70 & 1.31 \\
            \noalign{\smallskip}
            \hline
            \noalign{\smallskip}
            09 23:56 & 00:50 & 18.7 & 12288 & 2.00-1.61 & 1.53-1.64 \\
            10 01:38 & 02:04 &  8.9 &  4928 & 1.43-1.38 & 1.67 \\
            \noalign{\smallskip}
            \hline
      \end{tabular}
   \end{table}
 \end{center}
%

On Nov.\ 8, the asteroid was so bright (observatory-centric
distance was about 0.0021-0.0023\,AU) that it was detectable in
each frame. Sky frames -composed by averaging 7 frames taken
within 2\,sec- were subtracted. This successfully canceled out
sky variation similar to observations taken in chopping technique.
Aperture photometry with an aperture radius of 3 arcsec was
applied for each frame. Photometric values were determined by
averaging frames taken in a period of 5 minutes. Errors were
estimated as standard deviation of the photometric values.

On Nov.\ 9 the asteroid was already at about 0.0084-0.0089\,AU
distance and it was difficult to detect the asteroid on each frame.
Here, we added 92 frames into one image by shifting the frames to
compensate the asteroid's movement on the sky and subtracted the
averaged sky frames. The frame-to-frame shifts were calculated
from the ephemeris provided by the NASA Horizons web
page\footnote{JPL Horizons:
{\tt http://ssd.jpl.nasa.gov/horizons.cgi}}.
The asteroid images in the co-added frames appeared nearly point-like
and no noticeable extensions were detected. We applied aperture
photometry on the final sky-subtracted images. The final flux and
error values were again obtained by averaging all individual
photometric values.

In addition to the photometric error we also added a 5\% absolute
flux calibration error for the N-band data and a 7\% error
for the Q-band data based on the radiometric tolerance discussion
in Cohen et al.\ (\cite{cohen99}) and the information given in
the stellar template. These errors also include possible colour-terms
(estimated to be below 2\%) due to the different spectral shapes of
the star and the asteroid in the N- and Q-band filters.
In the first night (8/9 Nov.) $\alpha$~Tuc and the asteroid were
observed at similar airmass and similar PWV\footnote{Precipitable Water Vapour:
this is the main source of opacity at mid-infrared wavelengths.}-levels (based on APEX
measurements) and no additional corrections were needed. In the second
night (9/10 Nov.) 2005~YU$_{55}$ was observed in Q-band at a large
airmass close to 2.0 and a PWV of around 0.5\,mm, while $\alpha$~Tuc
was taken at an airmass of around 1.6 and a PWV-level of about 0.3\,mm.
Based on ATRAN model calculation of the atmospheric transmittance
vs.\ PWV for the 18\,$\mu$m filter we estimated that the derived Q-band
flux for 2005~YU$_{55}$ must be about 5-10\% too low. We increased the
derived 18.7\,$\mu$m flux of the second night by 8\% and gave a
10\% absolute flux calibration error (instead of 7\%) to compensate
for the additional source of uncertainty. The final calibrated flux
densities are given in Table~\ref{tbl:obsflx}.

\begin{table*}
     \caption{Observing geometries (miniTAO-centric) and final calibrated flux densities.
      Negative phase angles: after opposition (object was trailing the Sun). 
      An absolute flux calibration of 5\%
      (N-band) and 7/10\% (1$^{st}$/2$^{nd}$ day Q-band) has been added. The second day
      Q-band data point has been corrected for airmass/PWV effects (see text).}
     \label{tbl:obsflx}
     \begin{tabular}{lrrrrrrl}
        \noalign{\smallskip}
        \hline
        \hline
        \noalign{\smallskip}
Julian Date & $\lambda_{ref}$ & FD    & FD$_{err}$   & r$_{helio}$ & $\Delta_{obs}$ & $\alpha$ & Observatory/ \\
mid-time    & [$\mu$m]        & [Jy]  & [Jy]         &  [AU]      & [AU]            & [deg]    & Instrument \\
        \noalign{\smallskip}
        \hline
        \noalign{\smallskip}
2455874.46285 & 18.7 & 189.53 &  30.44 & 0.9904028 & 0.0021426484 & -97.17 & miniTAO/MAX38\tablefootmark{a} \\
2455874.46632 & 18.7 & 192.82 &  30.25 & 0.9904293 & 0.0021415786 & -96.46 & miniTAO/MAX38\tablefootmark{a} \\
2455874.46979 & 18.7 & 192.81 &  29.53 & 0.9904558 & 0.0021408565 & -95.74 & miniTAO/MAX38\tablefootmark{a} \\
2455874.47326 & 18.7 & 196.73 &  32.07 & 0.9904823 & 0.0021404827 & -95.03 & miniTAO/MAX38\tablefootmark{a} \\
2455874.47674 & 18.7 & 194.37 &  33.12 & 0.9905088 & 0.0021404572 & -94.32 & miniTAO/MAX38\tablefootmark{a} \\
2455874.48021 & 18.7 & 203.01 &  32.83 & 0.9905352 & 0.0021407803 & -93.61 & miniTAO/MAX38\tablefootmark{b} \\
2455874.48368 & 18.7 & 204.71 &  35.25 & 0.9905617 & 0.0021414518 & -92.89 & miniTAO/MAX38\tablefootmark{b} \\
2455874.48715 & 18.7 & 212.57 &  34.45 & 0.9905882 & 0.0021424716 & -92.18 & miniTAO/MAX38\tablefootmark{b} \\
2455874.49062 & 18.7 & 217.57 &  33.65 & 0.9906147 & 0.0021438391 & -91.47 & miniTAO/MAX38\tablefootmark{b} \\
2455874.49410 & 18.7 & 223.70 &  34.04 & 0.9906411 & 0.0021455542 & -90.76 & miniTAO/MAX38\tablefootmark{b} \\
2455874.49757 & 18.7 & 219.77 &  32.97 & 0.9906676 & 0.0021476157 & -90.05 & miniTAO/MAX38\tablefootmark{c} \\
2455874.50104 & 18.7 & 222.79 &  32.28 & 0.9906941 & 0.0021500229 & -89.34 & miniTAO/MAX38\tablefootmark{c} \\
2455874.50451 & 18.7 & 222.34 &  34.22 & 0.9907206 & 0.0021527749 & -88.63 & miniTAO/MAX38\tablefootmark{c} \\
2455874.50799 & 18.7 & 227.06 &  34.94 & 0.9907471 & 0.0021558704 & -87.93 & miniTAO/MAX38\tablefootmark{c} \\
2455874.51146 & 18.7 & 223.24 &  35.74 & 0.9907735 & 0.0021593083 & -87.22 & miniTAO/MAX38\tablefootmark{c} \\
2455874.52187 &  8.9 & 126.81 &  15.22 & 0.9908530 & 0.0021716598 & -85.13 & miniTAO/MAX38\tablefootmark{d} \\
2455874.52535 &  8.9 & 123.50 &  13.45 & 0.9908794 & 0.0021764507 & -84.43 & miniTAO/MAX38\tablefootmark{d} \\
2455874.52882 &  8.9 & 124.36 &  14.10 & 0.9909059 & 0.0021815751 & -83.74 & miniTAO/MAX38\tablefootmark{d} \\
2455874.54965 & 12.2 & 225.22 &  24.78 & 0.9910648 & 0.0022191984 & -79.67 & miniTAO/MAX38\tablefootmark{e} \\
2455874.55312 & 12.2 & 221.94 &  24.94 & 0.9910912 & 0.0022265916 & -79.00 & miniTAO/MAX38\tablefootmark{e} \\
2455874.55660 & 12.2 & 215.43 &  23.59 & 0.9911177 & 0.0022342979 & -78.34 & miniTAO/MAX38\tablefootmark{e} \\
2455874.56007 & 18.7 & 261.49 &  37.68 & 0.9911442 & 0.0022423147 & -77.68 & miniTAO/MAX38\tablefootmark{f} \\
2455874.56354 & 18.7 & 253.99 &  33.16 & 0.9911706 & 0.0022506389 & -77.03 & miniTAO/MAX38\tablefootmark{f} \\
2455874.56701 & 18.7 & 248.50 &  34.51 & 0.9911971 & 0.0022592671 & -76.38 & miniTAO/MAX38\tablefootmark{f} \\
2455874.57049 & 18.7 & 247.40 &  33.88 & 0.9912236 & 0.0022681963 & -75.74 & miniTAO/MAX38\tablefootmark{f} \\
2455874.57396 & 18.7 & 241.90 &  36.31 & 0.9912501 & 0.0022774233 & -75.10 & miniTAO/MAX38\tablefootmark{f} \\
2455874.57743 & 18.7 & 241.30 &  33.92 & 0.9912765 & 0.0022869445 & -74.47 & miniTAO/MAX38\tablefootmark{f} \\
2455875.51458 & 18.7 &  28.08 &   5.90 & 0.9983930 & 0.0084376885 & -19.23 & miniTAO/MAX38\tablefootmark{g} \\
2455875.57639 &  8.9 &  19.60 &   2.32 & 0.9988611 & 0.0089028990 & -18.57 & miniTAO/MAX38\tablefootmark{h} \\
\noalign{\smallskip}
\hline
     \end{tabular}
\tablefoot{\tablefoottext{a,b,c,d,e,f,g,h}{for the $\chi^2$ analysis in Section~\ref{sec:tpm} we used
           the mean fluxes of each group for calculation efficiency reasons.}}
\end{table*}

\subsection{Space far-infrared observations with Herschel-PACS}

\begin{figure}[h!tb]
 \rotatebox{0}{\resizebox{\hsize}{!}{\includegraphics{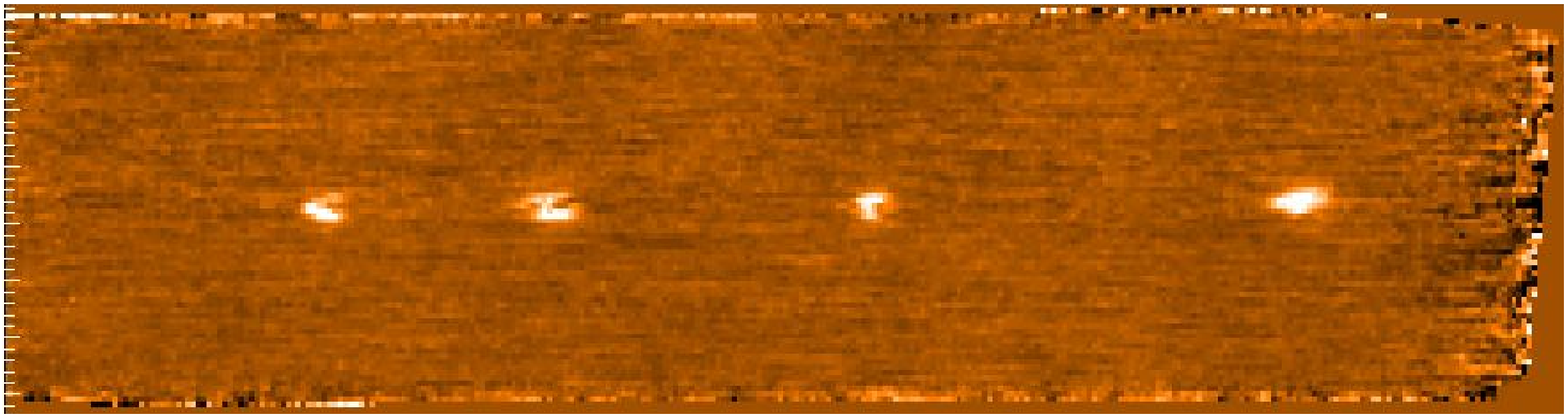}}}
 \rotatebox{0}{\resizebox{\hsize}{!}{\includegraphics{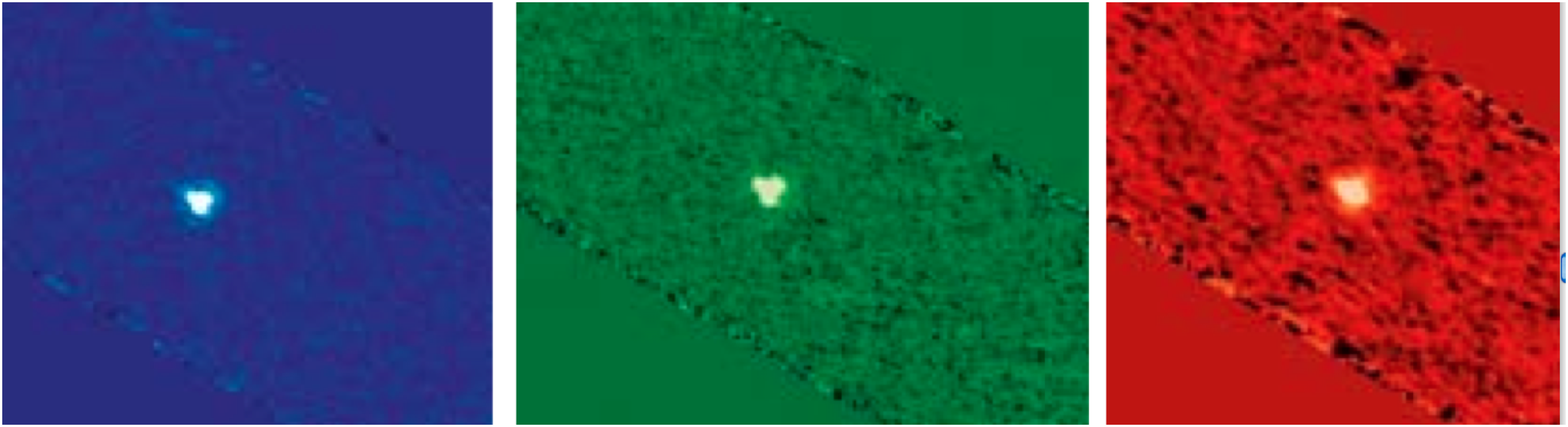}}}
  \caption{Top: Sky-projected PACS image of 2005~YU$_{55}$ at 70\,$\mu$m. Each of
           the 4 scan-legs has seen the target at a different position.
           Bottom: object-centered images of the target in the 3 filters:
                   blue (70\,$\mu$m), green (100\,$\mu$m), red (160\,$\mu$m).
                   The tripod-dominated point-spread-function is clearly
                   visible.
     \label{fig:pacs1}}
\end{figure}

The far-infrared observations with the Herschel space observatory
were reported by M\"uller et al.\ (\cite{mueller11b}).
2005~YU$_{55}$ crossed the entire visibility window ($\sim$60$^{\circ}$ to $\sim$115$^{\circ}$
solar elongation) in about 16\,hrs and its apparent motion was between 2.8 and
3.8\,$^{\circ}$/h, far outside the technical tracking limit of the 
satellite. Therefore, we performed two standard scan-map observations
of 240\,s length each -one in the 70/160\,$\mu$m (2011-Nov-10 14:52-14:56 UT,
OBSID 1342232729) and one in the 100/160\,$\mu$m filter combination (2011-Nov-10
14:57-15:01 UT, OBSID 1342232730)- at fixed times at pre-calculated positions on the sky.
Each scan-map consisted of 4 scan-legs of 14\,arcmin length and separated
by 4\,arcsec parallel to the apparent motion of the target and with a
scan-speed of 20$^{\prime \prime}$/s. During both scan-map observations
2005~YU$_{55}$ crossed the observed field-of-view and the target was
seen in each scan-leg. Figure~\ref{fig:pacs1} (top) shows the sky-projected
image of the 70\,$\mu$m band observations. The PACS photometer takes data
frames with 40\,Hz, but binned onboard by a factor of 4 before downlink.
We re-centered/stacked all frames where the satellite
was scanning with constant speed (about 1700 frames in each of the two
dual-band measurements) on the expected position of 2005~YU$_{55}$. The
results are shown in Fig.~\ref{fig:pacs1} (bottom). This technique
worked extremely well and one can clearly see many details of the
tripod-dominated point-spread-function. We performed aperture photometry
on the final calibrated images and estimated the flux error via photometry
on artificially implemented sources in the clean vicinity around
our target.
The fluxes were finally corrected for colour terms to obtain monochromatic
flux densities at the PACS reference wavelengths. These corrections are due to the differences
in spectral energy distribution between 2005~YU$_{55}$ and the assumed
constant energy spectrum $\nu$~F$_{\nu}$ = const.\ in the PACS
calibration scheme. The colour-corrections for objects in the temperature
range of $\approx$ 250 - 400\,K are 1.01, 1.03, 1.06 ($\pm$ 0.01)
in blue, green, red band respectively\footnote{PACS report PICC-ME-TN-038:\\ {\tt http://herschel.esac.esa.int/\-twiki/\-pub/\-Public/\-PacsCalibrationWeb/\-cc\_report\_v1.pdf}}.
The photometric error of the artificial sources were combined
quadratically with the absolute flux calibration errors (5\% in all 3 bands
based on the model uncertainties of the fiducial stars used in the
PACS photometer flux calibration scheme) and the error related to
the colour-correction (1\%).
The final monochromatic flux densities and their absolute
flux errors at the PACS reference wavelengths 70.0, 100.0 and 160.0\,$\mu$m
are listed in Table~\ref{tbl:obspacs}.

\begin{table*}
     \caption{Observing geometries (Herschel-centric) and final calibrated flux densities.
      Negative phase angles: after opposition.
     \label{tbl:obspacs}}
     \begin{tabular}{lrrrrrrl}
        \noalign{\smallskip}
        \hline
        \hline
        \noalign{\smallskip}
Julian Date & $\lambda_{ref}$ & FD    & FD$_{err}$   & r$_{helio}$ & $\Delta_{obs}$ & $\alpha$ & Observatory/ \\
mid-time    & [$\mu$m]        & [Jy]  & [Jy]         &  [AU]      & [AU]            & [deg]    & Instrument \\
        \noalign{\smallskip}
        \hline
        \noalign{\smallskip}
2455876.120565 &  70.0 & 12.35 & 0.63 &  1.002978 & 0.005403 & -70.88 &  Herschel-PACS \\
2455876.120565 & 160.0 &  2.55 & 0.13 &  1.002978 & 0.005403 & -70.88 &  Herschel-PACS \\
2455876.124075 & 100.0 &  6.87 & 0.35 &  1.003004 & 0.005415 & -70.62 &  Herschel-PACS \\
2455876.124075 & 160.0 &  2.66 & 0.14 &  1.003004 & 0.005415 & -70.62 &  Herschel-PACS \\
\noalign{\smallskip}
\hline
     \end{tabular}
\end{table*}


\subsection{Groundbased millimeter observations with the SMA}

\begin{table*}
     \caption{Observing geometries (SMA-centric) and final calibrated flux densities.
      Negative phase angles: after opposition.
     \label{tbl:obssma}}
     \begin{tabular}{lrrrrrrl}
        \noalign{\smallskip}
        \hline
        \hline
        \noalign{\smallskip}
Julian Date & $\lambda_{ref}$ & FD    & FD$_{err}$   & r$_{helio}$ & $\Delta_{obs}$ & $\alpha$ & Observatory/ \\
mid-time    & [$\mu$m]        & [Jy]  & [Jy]         &  [AU]      & [AU]            & [deg]    & Instrument \\
        \noalign{\smallskip}
        \hline
        \noalign{\smallskip}
2455874.95042 & 1328.9 & 0.075 &  0.020 & 0.9941165 & 0.0042883 & -34.66 & SMA/230\,GHz receiver \\
\noalign{\smallskip}
\hline
     \end{tabular}
\end{table*}

We performed observations of 2005~YU$_{55}$ a few hours past closest Earth
approach on Nov.\ 9, 2011 using the Submillimeter Array (SMA) located
near the summit of Mauna Kea in Hawaii. The SMA was operated in
separated sideband mode with 2\.GHz continuum bandwidth per sideband.
The lower sideband was tuned to 220.596\,GHz and the upper sideband
(USB) at 230.596\,GHz, providing a mean frequency of 225.596\,GHz, or
1328.9\,$\mu$m (covering the range from 1300.1 to 1359.0\,$\mu$m).  Complex
gains were obtained from several different quasars as the asteroid
moved across the sky.  The amplitude scale was corrected for Earth
atmospheric opacity through standard system temperature calibration,
and then corrected to the absolute (Jansky) scale by referencing to
observations of Uranus and Callisto, astronomical sources with flux
densities known to within $\sim$5\% at this frequency.

The measurements were difficult due to poor weather, particularly
atmospheric phase stability, and were further hampered by the
exceptionally rapid motion of the object. The asteroid's apparent
position at its fastest changed by $\sim$7$^{\prime \prime}$/s relative
to sidereal, which is significantly faster than the SMA phase tracking
system (the digital delay software, or DDS) was designed for.  To compensate, a
special version of the DDS was created which attempted to track the
phase on much shorter timescales.  However, this was only partly
successful and there were obvious signs of decorrelation (loss of
signal caused by the motion of the source relative to the tracked
phase center) on most baselines.  This required extensive data
flagging and secondary self-calibration of the amplitude, which
introduced significant systematic error in the flux density scale.

\begin{figure}[h!tb]
 \rotatebox{0}{\resizebox{\hsize}{!}{\includegraphics{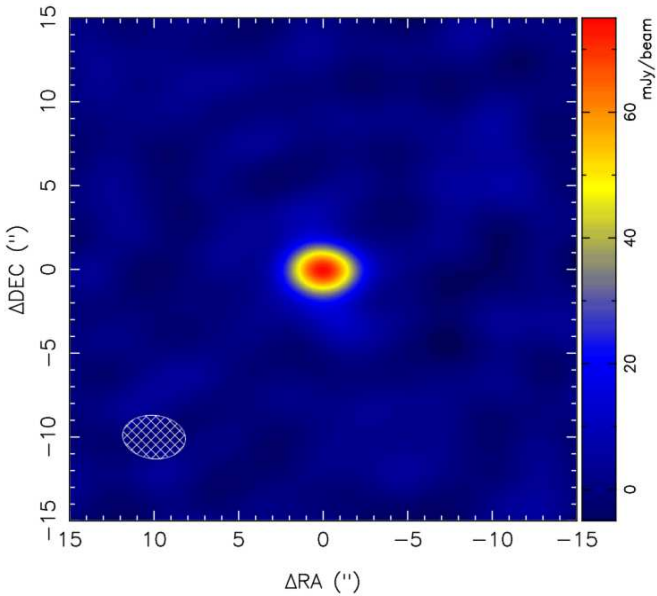}}}
  \caption{SMA image of 2005~YU$_{55}$ at 1.3\,mm. The ellipse represents
     the 2-dimensional full width at half maximum of the
     synthesized beam of the array, which is effectively the
     PSF of the instrument for the observations
     of an unresolved target.
     \label{fig:sma}}
\end{figure}

Despite these challenges, we obtained a clear detection of the
object. Figure~\ref{fig:sma} shows the 1.3\,mm image of 2005~YU$_{55}$ after both phase
reference calibration and further self-calibration.
The target itself was unresolved and the oblong image of the asteroid
in  Figure~\ref{fig:sma} is simply due to the PSF\footnote{Point Spread Function} of
the instrument for the observations, which is shown in the lower left corner as an
ellipse of 3.73$^{\prime \prime}$ $\times$ 2.58$^{\prime \prime}$ in size, with a
major axis position angle of 83.66$^{\circ}$ East of North.
Over the 3.5\,hours
of observation (UT Nov 9, 2011 09.16 - 12.46\,hrs) we further see the
expected drop in flux density as the source recedes, consistent with
the apparent size decrease with time.  While the detection is of high
signficance (SNR $\sim$35), the systematic problems of compensating for the
tracking-induced decorrelation along with the poor weather dominated
the flux-density error budget.  Taking all effects into account we
obtained a flux density of 75 $\pm$ 25\,mJy at observation mid-time (UT Nov
9, 2011 10:49, see Tbl.~\ref{tbl:obssma}).

\subsection{Auxiliary datasets}


\paragraph{Radar measurements.}

Nolan et al.\ (\cite{nolan10}), Busch et al.\ (\cite{busch12})
and Taylor et al.\ (\cite{taylor12a}; \cite{taylor12b})
presented results obtained by radar measurements using the
Arecibo S-band, the Deep Space Network Goldstone DSS-14 and DSS-13,
Green Bank Telescope and Arecibo/VLBA (radar speckle tracking).
They found 2005~YU$_{55}$ to be a dark (at radar/radio wavelengths),
spherical object of about 400\,m diameter and a rotation period of
roughly 18\,hrs (Nolan et al.\ \cite{nolan10}; Taylor et al.\ \cite{taylor12a}).
Busch et al.\ (\cite{busch12}) confirmed the nearly spheroidal
shape and determined the maximum dimensions of the object to be
360 $\pm$ 40\,m in all directions.
The radar team estimated the pole direction from the motion of the
radar speckle pattern during three days of observations after the flyby.
Combining the radar images and the speckle data excluded all prograde
pole directions, and restricted the possible retrograde poles to
($\lambda_{ecl}$, $\beta_{ecl}$) =
(20$^{\circ}$, -74$^{\circ}$) $\pm$ 20$^{\circ}$ with a rotation
period of 19.0 $\pm$ 0.5\,hrs and consistent with a principle-axis rotation.

\paragraph{Thermal infrared observations from Gemini-North/Michelle.}

Lim et al.\ (\cite{lim12a}; \cite{lim12b})
obtained thermal infrared photometry and spectroscopy in
N- and Q-band using the Michelle instrument at Gemini-North.
According to their thermal model analysis (T$_{ss}$ = 360 - 370\,K;
$\eta$ $\approx$ 1.25-1.5) the thermal measurements are
consistent with an object diameter of 400\,m, but the best
fit to their data was found for a size of 322 $\pm$ 18\,m
and a maximum subsolar temperature T$_{ss}$ of 409 $\pm$ 12\,K
(thermal model $\eta$ $\approx$ 0.93). More recently,
Lim et al.\ (\cite{lim12c}) combined their thermal data
with results from radar measurements and find now an
equatorial diameter of 380 $\pm$ 20\,m and a thermal
inertia $\Gamma$ $\approx$ 500 - 1500\,Jm$^{-2}$s$^{-0.5}$K$^{-1}$.
They also calculated values for the effective diameter
via radiometric techniques and based only on their thermal data.
They found an effective diameter of 310\,m for a low thermal inertia
of 350\,Jm$^{-2}$s$^{-0.5}$K$^{-1}$ and of 350\,m for a thermal
inertia of 1000\,Jm$^{-2}$s$^{-0.5}$K$^{-1}$ (DPS meeting \#44,
\#305.01 presentation).

\paragraph{Keck adaptive-optics (AO) imaging.}

Merline et al.\ (\cite{merline11}; \cite{merline12}) reported
on adaptive optics (AO) imaging of 2005~YU$_{55}$ during its
close fly-by on 2011 Nov 9 UT with the Keck~II AO system NIRC2.
The preliminary results were derived under the assumption of
a smooth triaxial ellipsoid having a principle-axis rotation
of 18\,hr. They found a preference for poles in the southern
sub-latitudes and an effective object diameter of 307 $\pm$ 15\,m.
This would be consistent with the radar-favoured retrograde
sense of rotation meaning that the object presented a warm
terminator during its close approach. In addition, they
give two explicit solutions: (a) prograde pole with
($\lambda_{ecl}$, $\beta_{ecl}$) = (339$^{\circ}$, +84$^{\circ}$) $\pm$ 6$^{\circ}$
and object dimensions of 337 $\times$ 324 $\times$ 267\,m ($\pm$ 15\,m in each dimension),
corresponding to a spherical equivalent diameter of 308 $\pm$ 9\,m;
(b) retrograde pole with
($\lambda_{ecl}$, $\beta_{ecl}$) = (22$^{\circ}$, -35$^{\circ}$) $\pm$ 15$^{\circ}$
and object dimensions of 328 $\times$ 312 $\times$ 245\,m ($\pm$ 15, 15, 30\,m)
corresponding to a spherical equivalent diameter of 293 $\pm$ 14\,m.

\paragraph{VLT-NACO speckle imaging observations.}

Sridharan et al.\ (\cite{sridharan12}) performed VLT-NACO speckle imaging
in Ks band in no-AO mode. The observations on 2005~YU$_{55}$ were
carried out one hour (10-min block) and two hours (15-min block)
after the closest Earth approach, interleaved by sky background and
calibration observations. The planned closed-loop AO observations failed
due to poor observing conditions and only no-AO mode (speckle imaging mode)
observations were possible. They found that 2005~YU$_{55}$ has a 
spherical shape with a mean diameter of about 270\,m. At the same time they
extracted a mean diameter of 261$\pm$20\,m $\times$ 310$\pm$30\,m from
edge-enhanced image reconstructions. The large uncertainties are due
to the theoretical resolution of 95\,m at the distance of the object
and the final image quality.

\paragraph{CCD photometric observations.}

CCD lightcurve measurements from different observers were
analysed by Warner et al.\ (\cite{warner12a}; \cite{warner12b}).
Their analysis resulted in two possible synodic periods
of 16.34 $\pm$ 0.01\,h with an amplitude of 0.24 $\pm$ 0.02\,mag
(9-17 Nov, 2011) and 19.31 $\pm$ 0.02\,h with an amplitude of
0.20 $\pm$ 0.02\,mag.
The first one was apparently supported by the initial radar
analysis, while the second one is now the currently favoured
solution by the radar team. The 19.31\,h lightcurve has a 
bimodal shape and there seem to be indications for a non-principal
axis rotation.
Due to a large phase angle coverage of the CCD data
they were also able to derive the absolute R-band magnitude
H$_R$ = 20.887 $\pm$ 0.042 and the phase slope parameter
G = -0.147 $\pm$ 0.014. With an assumed V-R value of 0.38
they calculated the absolute V-band magnitude
H$_V$ = 21.27 $\pm$ 0.05.

\paragraph{Absolute magnitude and phase curve.}

Based on Bessel R-band photometry and long-slit CCD spectrograms
during the 2010 and 2011 apparitions, Hicks et al.\ (\cite{hicks10};
\cite{hicks11}) reported an absolute R-band magnitude of
H$_R$ = 20.73 and a phase slope parameter G = -0.12 describing a
very steep phase curve which is typically found for low-albedo
C- and P-type asteroids. They measured a V-R colour of 0.37 mag
leading to an absolute V-band magnitude of H$_V$ = 21.1 $\pm$ 0.1.
An indpendent work by Bodewits et al.\ (\cite{bodewits11}) presented
a V-band absolute magnitude of H$_V$ = 21.2 when applying a phase
curve derived from UV measurements (G$_{UV}$ = -0.13).

%

\section{Thermophysical model analysis}
\label{sec:tpm}

For the analysis of our thermal data (miniTAO/MAX38, SMA, Herschel/PACS)
we applied a thermophysical model (TPM) which is based on the work by
Lagerros (\cite{lagerros96}; \cite{lagerros97}; \cite{lagerros98}).
This model is frequently and successfully applied to near-Earth
asteroids (e.g., M\"uller et al.\ \cite{mueller04b};
M\"uller et al.\ \cite{mueller05};
M\"uller et al.\ \cite{mueller11a};
M\"uller et al.\ \cite{mueller12}),
to main-belt asteroids (e.g., M\"uller \& Lagerros \cite{mueller98}; 
M\"uller \& Blommaert \cite{mueller04a}), and also to more distant
objects (e.g.\ Horner et al.\ \cite{horner12}; Lim et al.\ \cite{lim10}).
The TPM takes into account the true observing and illumination geometry
for each observational data point, a crucial aspect for the interpretation
of our 2005~YU$_{55}$ observations which cover a wide range of phase angles.
The TPM allows to specify a shape model
and spin-vector properties. The heat conduction into the surface is
controlled by the thermal inertia $\Gamma$. The observed mid- and far-IR
fluxes are connected to the hottest regions on the asteroid surface and 
dominated by the diurnal heat wave. The seasonal heat wave is less important
and therefore not considered here. The infrared beaming effects are calculated
via a surface roughness model, described by segments of hemispherical craters.
Here, mutual heating is included and the true crater illumination and
the visibility of shadows is considered.
The level of roughness is driven by the r.m.s.\ of the surface slopes which
correspond to a given crater depth-to-radius value combined with the fraction
of the surface covered by craters, see also Lagerros (\cite{lagerros96}) for
further details. We used a constant emissivity of 0.9 at all wavelengths,
knowing that the emissivity can decrease beyond $\sim$200\,$\mu$m in some
cases (e.g., M\"uller \& Lagerros \cite{mueller98}; \cite{mueller02}). All of our data -except the
SMA data point which has a large errorbar- have been taken at wavelength $<$200\,$\mu$m
and the constant emissivity is therefore a valid assumption.
The TPM input parameters and applied variations are listed in Table.~\ref{tbl:tpm_params}.

  \begin{table}[h!tb]
    \begin{center}
    \caption{Summary of general TPM input parameters and applied ranges.
             \label{tbl:tpm_params}}
    \begin{tabular}{lcl}
      \hline
      \hline
      \noalign{\smallskip}
      Param.\  &  Value/Range & Remarks \\
      \noalign{\smallskip}
      \hline
      \noalign{\smallskip}
    $\Gamma$            & 0...3000                  & J\,m$^{-2}$\,s$^{-0.5}$\,K$^{-1}$, thermal inertia \\
                        &                           & (25 values spread in log-space) \\
    $\rho$              & 0.1...0.8                 & r.m.s. of surface slopes, steps of 0.1 \\
    $f$                 & 0.6\tablefootmark{a}      & surface frac.\ covered by craters \\
    $\epsilon$          & 0.9\tablefootmark{b}      & $\lambda$-independent emissivity \\
    $H_{\rm{V}}$-mag.\  & 21.2\,$\pm$\,0.15\,mag       & average of published values \\
    G-slope             & -0.13\,$\pm$\,0.02            & average of published values \\
    shape               & spherical/ellipsoidal         & info from radar and AO \\
    P$_{\mathrm{sid}}$ [h]       & 16.34\,h; 19.31\,h   & Warner et al.\ (\cite{warner12a}; \cite{warner12b}) \\
      \noalign{\smallskip}
      \hline
      \noalign{\smallskip}
    spin-axis                       &  (20.0$^{\circ}$, -74.0$^{\circ}$) $\pm$ 20$^{\circ}$ & Busch et al.\ (\cite{busch12}) \\                              
    ($\lambda_{ecl}$,$\beta_{ecl}$) & (339.0$^{\circ}$, +84.0$^{\circ}$) $\pm$ 6$^{\circ}$  & Merline et al.\ (\cite{merline11}; \cite{merline12}) \\        
                                    &  (22.0$^{\circ}$, -35.0$^{\circ}$) $\pm$ 15$^{\circ}$ & Merline et al.\ (\cite{merline11}; \cite{merline12}) \\        
                                    & (309.3$^{\circ}$, +89.5$^{\circ}$)\tablefootmark{c}   & obliquity 0$^{\circ}$ (prograde) \\                            
                                    & (129.3$^{\circ}$, -89.5$^{\circ}$)\tablefootmark{d}   & obliquity 180$^{\circ}$ (retrograde) \\                        
                                    & (337.2$^{\circ}$, -13.9$^{\circ}$)                    & pole-on case1 for Herschel obs.\ \\                         
                                    & (157.2$^{\circ}$, +13.9$^{\circ}$)                    & pole-on case2 for Herschel obs.\ \\                         
                                    & (273.0$^{\circ}$,  +1.7$^{\circ}$)                    & pole-on case1 for TAO/MAX38  \\                            
                                    &  (93.0$^{\circ}$,  -1.7$^{\circ}$)                    & pole-on case2 for TAO/MAX38  \\                            
                                    & (337.2$^{\circ}$, +76.1$^{\circ}$)                    & equ.-on case1 for Herschel obs.\ \\                      
                                    & (157.2$^{\circ}$, -76.1$^{\circ}$)                    & equ.-on case2 for Herschel obs.\ \\                      
                                    & (273.0$^{\circ}$, -88.3$^{\circ}$)                    & equ.-on case1 for TAO/MAX38  \\                         
                                    &  (93.0$^{\circ}$, +88.3$^{\circ}$)                    & equ.-on case2 for TAO/MAX38  \\                         
                                    & (0/90/180/270$^{\circ}$, $\pm$60$^{\circ}$)           & intermediate orientations \\                                   
                                    & (0/90/180/270$^{\circ}$, $\pm$30$^{\circ}$)           & intermediate orientations \\                                   
                                    & (0/90/180/270$^{\circ}$, 0$^{\circ}$)                 & pole in ecliptic plane \\                                      
     \noalign{\smallskip}
     \hline
     \noalign{\smallskip}
    \end{tabular}
\tablefoot{\tablefoottext{a}{see Lagerros \cite{lagerros98} section 3.3};
           \tablefoottext{b}{see text for further details};
           \tablefoottext{c}{spin-axis orientation close to ecliptic north pole};
           \tablefoottext{d}{spin-axis orientation close to ecliptic south pole}}
    \end{center}
  \end{table}

\subsection{Using a spherical shape model}

\begin{figure}[h!tb]
 \rotatebox{90}{\resizebox{!}{\hsize}{\includegraphics{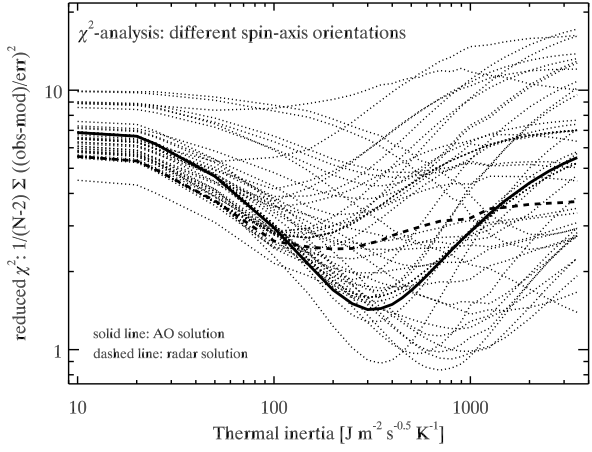}}}
  \caption{Calculation of reduced $\chi^2$-values for all specified spin-axis
   orientations, a fixed rotation period of 19.31\,h and an intermediate
   surface roughness level (r.m.s.\ of surface slopes of 0.3). The prograde AO solution (solid line) and the
   radar solution (dashed line) for the spin-vector are indicated in the figure.
     \label{fig:chi2_sv}}
\end{figure}

We started our analysis with a spherical shape model to see which spin-axis
orientations, sizes, geometric albedos, and thermal properties produce
acceptable solutions with reduced $\chi^2$-values\footnote{reduced $\chi^2$-values were
calculated via $\chi^2_{reduced}$ = 1/(N-$\nu$) $\sum$ ((obs-mod)/err)$^2$, with $\nu$ being the
number of free degrees of freedom; here $\nu$=2 since we solve for diameter and thermal inertia;
{\it obs} is the observed and {\it mod} the model flux, {\it err} the absolute photometric error.}
around or below 1.0. For the spin-axis solutions we used all values specified in literature
and many additional orientations to cover the entire $\lambda_{ecl}$-$\beta_{ecl}$ space.
For the calculation of the reduced $\chi^2$-curves we consider the true observing and
illumination constellation (helio-centric and observer-centric distances, phase angle,
spin-axis orientation) for each epoch  and then we compare
with the corresponding measurement. These calculations are done for a wide range
of thermal inertias and different levels of surface roughness as specified in
Table~\ref{tbl:tpm_params}. An example for the application of this technique
can be found in M\"uller et al.\ (\cite{mueller11a}). Each model setup produces
a curve of reduced $\chi^2$-values as a function of thermal inertia. Figure~\ref{fig:chi2_sv}
shows these curves for all different spin-axis orientation, a rotation period of
19.31\,h and an intermediate level of surface roughness (r.m.s.\ of surface slopes of 0.3).
Reduced $\chi^2$-values around or below 1.0 correspond to TPM solutions
which explain all observed fluxes in a statistically acceptable way. There are several
spin-axis orientations which produce an excellent match to all our thermal
measurements at thermal inertia values in the range between approximately 200 and
1500\,Jm$^{-2}$s$^{-0.5}$K$^{-1}$.

\begin{figure}[h!tb]
 \rotatebox{0}{\resizebox{\hsize}{!}{\includegraphics{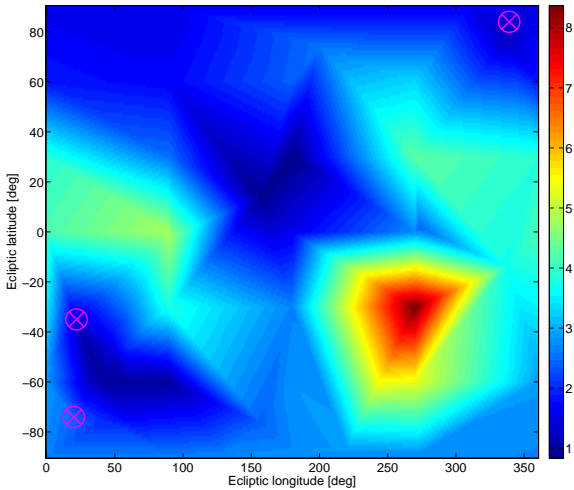}}}
  \caption{The $\chi^2$-minima calculated for all spin-axis orientations 
           listed in Table~\ref{tbl:tpm_params} and for an intermediate level of
           roughness (r.m.s.\-slope 0.3). The dark blue zones indicate
           spin-poles which allow us to obtain an acceptable match to
           all thermal data simultaneously (reduced $\chi^2$-values around
	   or below 1.0). The radar and both AO solutions are indicated by
	   the crossed circles. Note that the size, albedo and thermal inertia
	   are free parameters and only the best possible solution for each spin-axis
	   has been considered. \label{fig:chi2_0306}}
\end{figure}

The distribution of the reduced $\chi^2$-minima along the ecliptic
longitudes and latitudes is shown in Figure~\ref{fig:chi2_0306}. 
There are large zones in the $\lambda_{ecl.}$-$\beta_{ecl}$-space
which can be excluded with high probability (light blue, green, yellow, red zones),
but there remain several possible spin-axis orientations compatible
with our dataset (dark blue zones), including
the radar and AO solutions.

Both figures (Figs.~\ref{fig:chi2_sv} \& \ref{fig:chi2_0306})
have a slight dependency on the selected surface roughness
(for both figures we have used r.m.s.\ of surface slopes of 0.3).
In general, lower roughness (r.m.s.\ of surface slopes at 0.1) produces lower
$\chi^2$-minima and at smaller thermal inertia values going down
to about 200\,Jm$^{-2}$s$^{-0.5}$K$^{-1}$. Higher values for the
surface roughness (r.m.s.\ of surface slopes of $\ge$0.5) shift the
$\chi^2$-minima to values well above 1.0 and towards higher
thermal inertia going up to about 1500\,Jm$^{-2}$s$^{-0.5}$K$^{-1}$.
It is interesting to note that the prograde AO solution (solid line
in Fig.~\ref{fig:chi2_sv}) works very well ($\chi^2$-minima very
close to 1.0) for a low surface roughness, while the radar solution
produces a better match in case of a high surface roughness.

\subsection{Influence of the spin-axis orientation}

\begin{figure}[h!tb]
 \rotatebox{90}{\resizebox{!}{\hsize}{\includegraphics{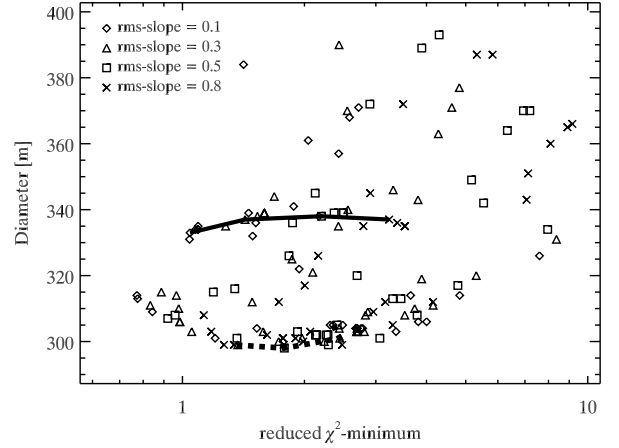}}}
 \rotatebox{90}{\resizebox{!}{\hsize}{\includegraphics{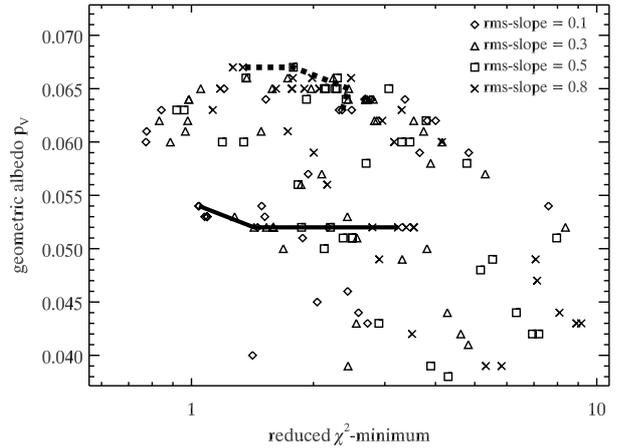}}}
  \caption{The distribution of the $\chi^2$-minima and the related
           effective diameter (top) and geometric albedos (bottom). The four different levels of roughness are
           indicated by different symbols. The values for the prograde AO
           solution (solid line) and the radar solution (dashed line)
           are connected in the figures.
     \label{fig:size_pv}}
\end{figure}

As a next step, we investigate the influence of different spin-axis
orientations on the size and albedo solutions. We determined the
$\chi^2$-minima for all listed spin-axis orientations and for four
different levels of roughness (r.m.s.\ of surface slopes at 0.1, 0.3,
0.5 and 0.8). Figure~\ref{fig:size_pv} shows how the corresponding
radiometric sizes and geometric albedos are
distributed in the reduced $\chi^2$-picture. We connected the four
$\chi^2$-minima belonging to the AO-solution (solid line) and the ones
belonging to the radar solution (dashed line) in Fig.~\ref{fig:size_pv}.
These lines show that the connected size and albedo values remain
stable, just the fit gets better (lower $\chi^2$-minima) for specific
roughness settings. We also found that the derived thermal inertias
change significantly with roughness at similar $\chi^2$-values, indicating
that we cannot resolve the degeneracy between roughness and thermal
inertia with our dataset. A smoother surface is connected
to lower values for the thermal inertia, the rougher surfaces require
higher thermal inertias. 

\begin{figure}[h!tb]
 \rotatebox{0}{\resizebox{\hsize}{!}{\includegraphics*{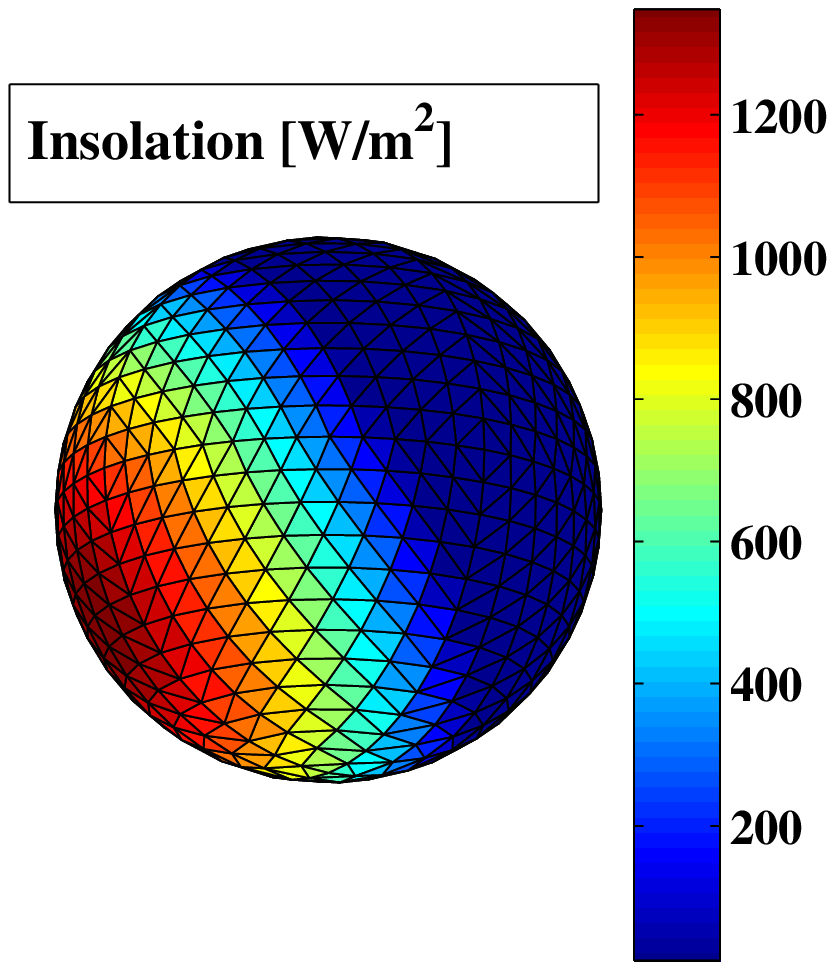}}}
 \rotatebox{0}{\resizebox{\hsize}{!}{\includegraphics*{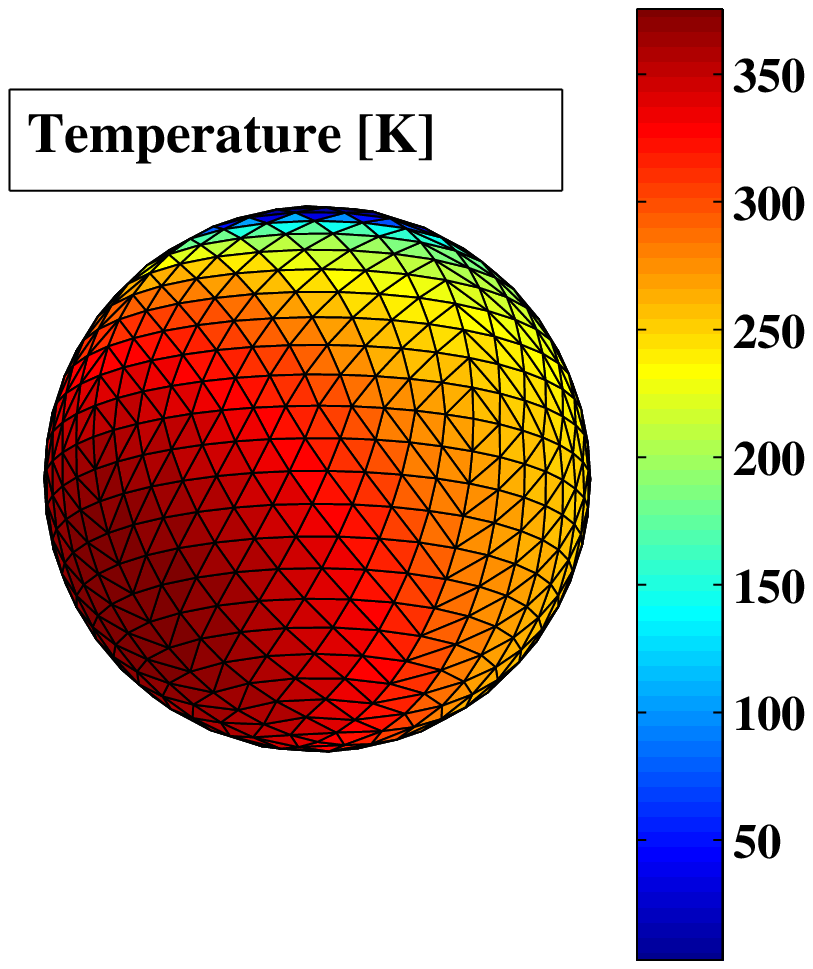}}}
  \caption{TPM picture of 2005~YU$_{55}$ as seen from Herschel on 2011-Nov-10 14:55 UT in the
           object-centered reference frame (z-axis along the object's rotation axis) and with
           the Sun at a phase angle of -71$^{\circ}$, spin-axis orientation: ($\lambda_{ecl}$, $\beta_{ecl}$)
	   = (60$^{\circ}$, -60$^{\circ}$), spherical shape model with a total of 800 facets. Top: insolation in W/m$^2$. Bottom:
           temperature in K.
     \label{fig:tpmfig}}
\end{figure}

The thermal data are compatible with different spin-axis orientations,
but the size, the geometric albedo and also the possible thermal
inertias are very well constrained by our thermal dataset. The best
solutions are found for an effective diameter of about 310\,m, if we 
include the best solutions for the prograde AO spin-axis and the
radar spin-axis orientations, then the possible diameter range goes
from 295 to 335\,m (see Fig.~\ref{fig:size_pv}, top). For the
geometric albedo we find a value of about 0.062 and a possible
range between 0.053 to 0.067 (see Fig.~\ref{fig:size_pv}, bottom).
Figure~\ref{fig:tpmfig} shows how our best TPM solution translates
the insolation during the epoch of the Herschel measurement
into a thermal picture of the surface as seen from Herschel.
For the calculations we used a spin-axis orientation of
($\lambda_{ecl}$, $\beta_{ecl}$) = (60$^{\circ}$, -60$^{\circ}$) and
a spherical shape model with a total of 800 facets.
The large influence of the thermal inertia in combination with the
object's rotation is the reason for the warm temperatures also in
regions without direct illumination.

\section{Discussions}
\label{sec:dis}

\subsection{Comparison with the radar results}

The comparison between the radar results (Busch et al.\ \cite{busch12})
and our findings is very interesting. If we use the radar diameter
(360 $\pm$ 40\,m, close to a spheroidal shape) and the spin-axis properties
([$\lambda_{ecl}$, $\beta_{ecl}$] = [20$^{\circ}$, -74$^{\circ}$] $\pm$ 20$^{\circ}$,
P$_{sid}$ = 19.0 $\pm$ 0.5\,h) it is not possible to find an acceptable
match to our thermal measurements. The reduced $\chi^2$-minima
stay always well above 2.0 and the match between TPM-predictions and observed
fluxes is very poor. Even for the lowest diameter limit of 320\,m
the model calculations would exceed the measured fluxes systematically by 15-25\%.
At a diameter of 360\,m the model fluxes are already 30-40\% above the
measurements. The radar size estimates are -as the radiometric size estimates-
model dependent. The spin-axis orientation as well as the rotation rate
have a larger influence on the radar solution (e.g., Ostro et al.\ \cite{ostro02})
than they have on the radiometric
solution. The radar images are dominated by the surface part which is closest 
to the antenna while the thermal data are tidely connected to the entire
cross-section at the moment of observation. This might explain the differences
between both techniques.

However, we do find an acceptable match to all thermal data if
we just use the radar spin-properties combined with a high level
of surface roughness (r.m.s.\ of surface slopes of 0.8). But the
corresponding diameter is only 299\,m -well outside the radar derived
range- with a p$_V$=0.067 and a thermal inertia of
400\,Jm$^{-2}$s$^{-0.5}$K$^{-1}$. In fact, all high obliquity cases
with $\beta_{ecl}$ $\le$ -60$^{\circ}$ (retrograde sense of rotation)
produce small diameters in the range 300-310\,m, while only the low
obliquity cases with $\beta_{ecl}$ $\ge$ +60$^{\circ}$ (prograde sense
of rotation) produce larger effective diameters of 325-340\,m.

\subsection{Comparison with AO and speckle results}

The Keck AO results presented by Merline et al.\ (\cite{merline12}) compare
better with our findings. Table~\ref{tbl:ao_comp} summarises the AO and
our radiometric results.

\begin{center}
\begin{table}[h!tb]
\caption{Comparison between AO results and our findings.}
\label{tbl:ao_comp}
\begin{tabular}{l c c c l}
\noalign{\smallskip}
\hline\hline
\noalign{\smallskip}
sense of  & spin-axis                        & AO-size & AO D$_{equ}$ & TPM-D$_{equ}$ \\
rotation  & ($\lambda_{ecl}$, $\beta_{ecl}$) & [m]     & [m]          & [m] \\
\noalign{\smallskip}
\hline
\noalign{\smallskip}
prograde   &  339$^{\circ}$, +84$^{\circ}$ & 337$\times$324$\times$267 & 308 $\pm$ 9 & 333 \\
retrograde &   22$^{\circ}$, -35$^{\circ}$ & 328$\times$312$\times$245 & 293 $\pm$ 14 & 299\tablefootmark{a} \\
retrograde & \multicolumn{2}{l}{southern poles} & 307 $\pm$ 15 & 300-310\tablefootmark{b} \\
\noalign{\smallskip}
\hline
\end{tabular}
\tablefoot{\tablefoottext{a}{this solution requires an unacceptably high thermal inertia
                             of well above 2000\,Jm$^{-2}$s$^{-0.5}$K$^{-1}$; }
           \tablefoottext{b}{diameter range of all high obliquity cases $\beta_{ecl}$ $\le$ -60$^{\circ}$}}
\end{table}
\end{center}

The southern rotational poles are not specified in detail by Merline et al.\ (\cite{merline12}),
but here we see for the first time an agreement between the derived sizes. The originally
specified retrograde pole towards an ecliptic latitude of -35$^{\circ}$ is very 
unlikely: acceptable TPM solutions (with reduced $\chi^2$-minima below 2.0) are only
found if the thermal inertia would be well above 2000\,Jm$^{-2}$s$^{-0.5}$K$^{-1}$,
an unrealistically high value which has never been measured before. It should be
noted here that the highest derived thermal inertias are still below 1000\,Jm$^{-2}$s$^{-0.5}$K$^{-1}$ (e.g., Delbo
et al.\ \cite{delbo07}) and that our mid- to far-IR data originate in the top layer
on the surface. We don't see any signatures of sub-surface layers where the 
thermal inertia could be significantly higher (Keihm et al.\ \cite{keihm12}).

The speckle observation in no-AO mode presented by Sridharan et al.\ (\cite{sridharan12})
revealed a roughly spheroidal shape with a mean diameter of 270\,m. By using a
more sophisticated reconstruction technique they estimated a mean diameter of 261 $\pm$ 20\,m
$\times$ 310 $\pm$ 30\,m, corresponding to an object-averaged size of approximately
285 $\pm$ 25\,m. Within the errorbars, this value agrees with our radiometrically
derived diameter of 300-310\,m and it also creates doubts if the large radar size
is realistic. The indications for a diameter close to 300\,m makes also
the various prograde solutions more unlikely, which all require diameters
in the range 325-340\,m.


\subsection{Spin-axis properties}

Combining the spin-axis information given by Busch et al.\ (\cite{busch12}),
Merline et al.\ (\cite{merline12}) and our findings (see Fig.~\ref{fig:chi2_sv}),
our analysis supports a retrograde sense of rotation with a possible spin-axis orientation
of ($\lambda_{ecl}$, $\beta_{ecl}$) = (60$^{\circ}$ $\pm$ 30$^{\circ}$, -60$^{\circ}$ $\pm$ 15$^{\circ}$).
The relatively large errors in ($\lambda_{ecl}$, $\beta_{ecl}$) are covering also the
possible solutions connected to the different roughness levels mentioned before.
If we use this solution, then the size estimate from AO observations matches our radiometrically
derived optimal size and we also have an agreement with the radar derived spin-pole.
The discrepancy with the radar size remains.

Our thermal observations cover a wide range of phase angles,
wavelengths and different illumination and observing geometries.
This allowed us to exclude many spin-axis orientations
(see Fig.~\ref{fig:chi2_0306}).
Nevertheless, we could not find a strong preference for a single
spin-axis orientation nor for the sense of rotation. Even very
extreme solutions like the retrograde radar solution and the
prograde AO solution seem to explain the data equally well.
This is very surprising. Based on our previous modeling experiences
for 1999~JU$_3$ (M\"uller et al.\ \cite{mueller11a}) and 1999~RQ$_{36}$
(M\"uller et al.\ \cite{mueller12}) based on much smaller sets of
thermal data, we expected to find a unique spin-axis solution.
But this might be an indication that 2005~YU$_{55}$ is a tumbler
with a strongly time-dependent orientation of the spin-axis
(for further details on tumbling asteroids see Pravec et al.\
\cite{pravec05}).
Busch et al.\ (\cite{busch12}) speculated already about the
possibility that terrestrial tides might have torqued the
object into a non-principal axis spin state. However, their
observations are consistent with a principle-axis rotation.
Warner et al.\ (\cite{warner12b}) found two, non-commensurate
solutions for the rotation period (16.34 $\pm$ 0.01\,h; 
19.31 $\pm$ 0.02\,h) which they could not fully explain. They
suggest that a non-principal axis rotation should be considered.
After the radar and lightcurve analysis, the thermal analysis
is now also pointing towards the possibility of a non-principle
axis rotation.

We also looked into the influence of the two published 
rotation periods. But the $\approx$3\,h difference between
the two available periods did not affect our radiometric solutions
significantly. The longer rotation period
is typically requiring slightly higher inertias to produce
the same disk-integrated flux, but this is a marginal effect
here in this case.

\subsection{Comparison with other thermal measurements}

Instead of comparing our TPM radiometric results with the 
preliminary results produced by Lim et al.\ (\cite{lim12a}; \cite{lim12b}; \cite{lim12c})
via a simple thermal model, we predicted flux densities for the 
epochs and the wavelength bands of the Michelle/Gemini North 
observations shown in Figure~2 in Lim et al.\ ({\cite{lim12b}).
For the TPM prediction we simply used our best effective diameter (310\,m)
and albedo (p$_V$ = 0.062) solution connected to our preferred
spin-axis orientation of ($\lambda_{ecl}$, $\beta_{ecl}$) = (60$^{\circ}$, -60$^{\circ}$).
The thermal inertia and roughness levels are less well constrained
and our dataset does not allow to break the degeneracy between these
two parameters. A low roughness (r.m.s.\ of surface slopes = 0.1) combined with
small values of the thermal inertia of about 200\,Jm$^{-2}$s$^{-0.5}$K$^{-1}$ would
explain our measurements as well as higher roughness levels
(r.m.s.\ of surface slopes = 0.5) combined with higher thermal inertia around
800\,Jm$^{-2}$s$^{-0.5}$K$^{-1}$.
We selected an intermediate solution (r.m.s.\ of surface slopes = 0.3;
thermal inertia = 500\,Jm$^{-2}$s$^{-0.5}$K$^{-1}$).

The Gemini-North/Michelle photometry shown in the Figure~2 in Lim et al.\ 
(\cite{lim12b}) was taken on 09-Nov-2011 11:02-11:15 UT
($\alpha$ = -34.0$^{\circ}$, r = 0.994\,AU, $\Delta$ = 0.004\,AU)
and on 10-Nov-2011 09:32 - 11:52 UT 
($\alpha$ = -15.5$^{\circ}$, r = 1.001\,AU, $\Delta$ = 0.012\,AU).
Since the calibrated flux densities and errors are not explicitly given,
we only could do a qualitative comparison.
Table~\ref{tbl:michelle_comp} shows our TPM prediction for both
epochs and the Michelle reference wavelengths in Jansky and W/m$^2$/$\mu$m.

\begin{center}
\begin{table}[h!tb]
\caption{TPM flux predictions for the Michelle bands and both
  observing epochs.}
\label{tbl:michelle_comp}
\begin{tabular}{rrcrc}
\noalign{\smallskip}
\hline\hline
\noalign{\smallskip}
Wavelength  & \multicolumn{2}{c}{09-Nov-2011 11:08UT} & \multicolumn{2}{c}{10-Nov-2011 10:50UT} \\
 $\lambda_{c}$ [$\mu$m]   &   [Jy] & [W/m$^2$/$\mu$m] & [Jy] & [W/m$^2$/$\mu$m] \\ 
\noalign{\smallskip}
\hline
\noalign{\smallskip}
 7.9 &  69.2 & 3.3e-12 & 10.6 & 5.1e-13 \\	
 8.8 &  83.7 & 3.2e-12 & 12.9 & 5.0e-13 \\	
 9.7 &  95.3 & 3.0e-12 & 14.7 & 4.7e-13 \\	
10.3 & 101.5 & 2.9e-12 & 15.6 & 4.4e-13 \\	
11.6 & 110.8 & 2.5e-12 & 17.0 & 3.8e-13 \\	
12.5 & 114.6 & 2.2e-12 & 17.6 & 3.4e-13 \\	
18.5 & 109.2 & 1.0e-12 & 16.6 & 1.5e-13 \\	
\noalign{\smallskip}
\hline
\end{tabular}
\end{table}
\end{center}

Our TPM-predictions agree very well with the
observed fluxes and errorbars presented in Lim et
al.\ (\cite{lim12b}). For the first epoch we estimated that the agreement
is within about 10\% at all wavelengths, while for the
second epoch the TPM prediction seems to be about 5-15\% below the
observed fluxes.

We also tested the low-roughness/low-inertia case mentioned before and indeed
it produces very similar fluxes and the agreement is on a similar level.
The high-roughness/high-inertia case is less convincing, the TPM predictions are
systematically low by 5-20\%. The Michelle/Gemini North data favour a thermal inertia
value in the range 200-700\,Jm$^{-2}$s$^{-0.5}$K$^{-1}$, combined with an intermediate to low
roughness level (r.m.s.\ of surface slopes 0.1-0.5), also in agreement with the
lowest reduced $\chi^2$-values in Fig.~\ref{fig:chi2_sv}.

\subsection{Overall fit to the measurements}



We tested the quality of the final solution for 2005~YU$_{55}$ against
the observed and calibrated flux densities by calculating the TPM predictions
for each of data point listed in Tables~\ref{tbl:obsflx}, \ref{tbl:obspacs}, and
\ref{tbl:obssma}. The observed and calibrated mono-chromatic flux densities are
shown in Fig.~\ref{fig:sed} together with the TPM predictions for the specific
observing geometries. The observation/TPM ratios are very sensitive to wavelength-dependent
effects (related surface roughness and thermal inertia), phase-angle dependent
effects (a wrong thermal inertia would cause before/after opposition
asymmetries), and shape effects (ratios as a function of rotational phase).
An overall ratio close to 1.0 indicates that the size and thermal properties
(and in second order also albedo) are correctly estimated.
Figure~\ref{fig:obsmod} shows how well our final TPM solution explains
our thermal data covering a wide range of wavelengths from 8.9\,$\mu$m to 1.3\,mm
and taken at very different phase angles ranging from -97$^{\circ}$ to -18$^{\circ}$.
No trends with wavelength nor with phase angle can be seen.

\begin{figure}[h!tb]
 \rotatebox{90}{\resizebox{!}{\hsize}{\includegraphics{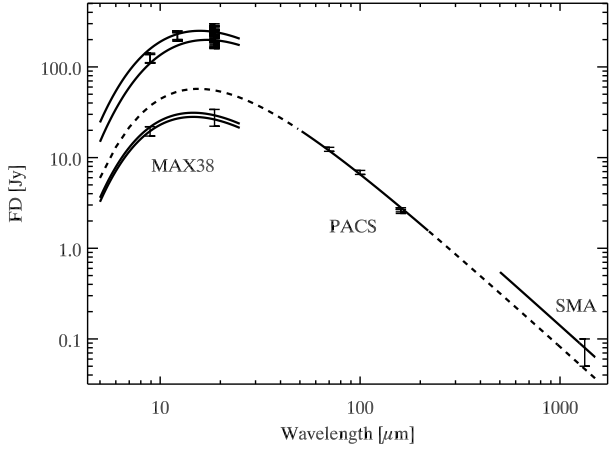}}}
  \caption{Observered and calibrated flux densities together with the corresponding TPM prediction. The model
           predictions for the MAX38 data are shown at the start and end time of each observing day. The distance
	   between observer and target and also the phase angle were rapidly changing during the close encounter
	   period of three days. For the PACS data the model prediction from 5 to 1500\,$\mu$m is shown.
     \label{fig:sed}}
\end{figure}

\begin{figure}[h!tb]
 \rotatebox{90}{\resizebox{!}{10cm}{\includegraphics{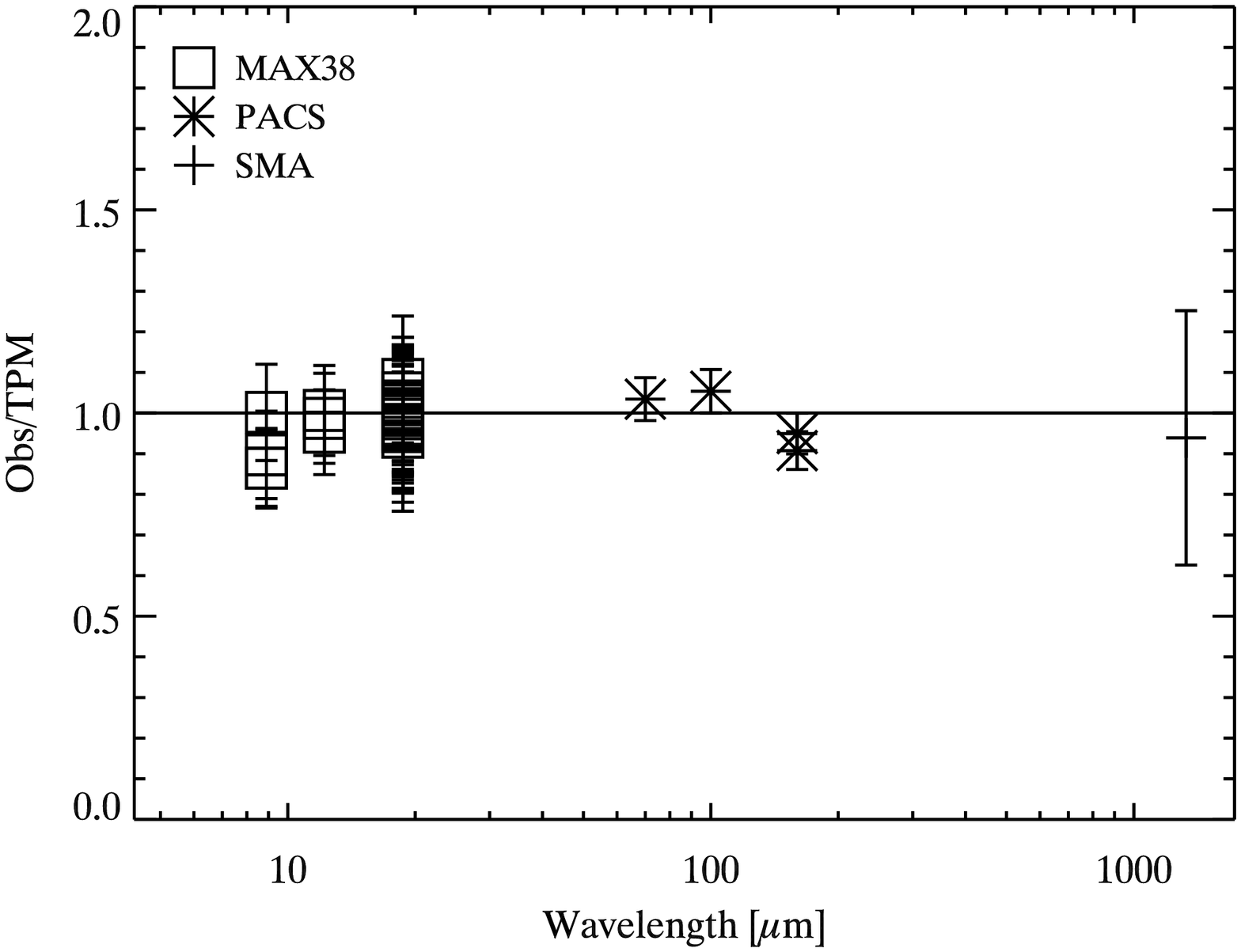}}}
 \rotatebox{90}{\resizebox{!}{10cm}{\includegraphics{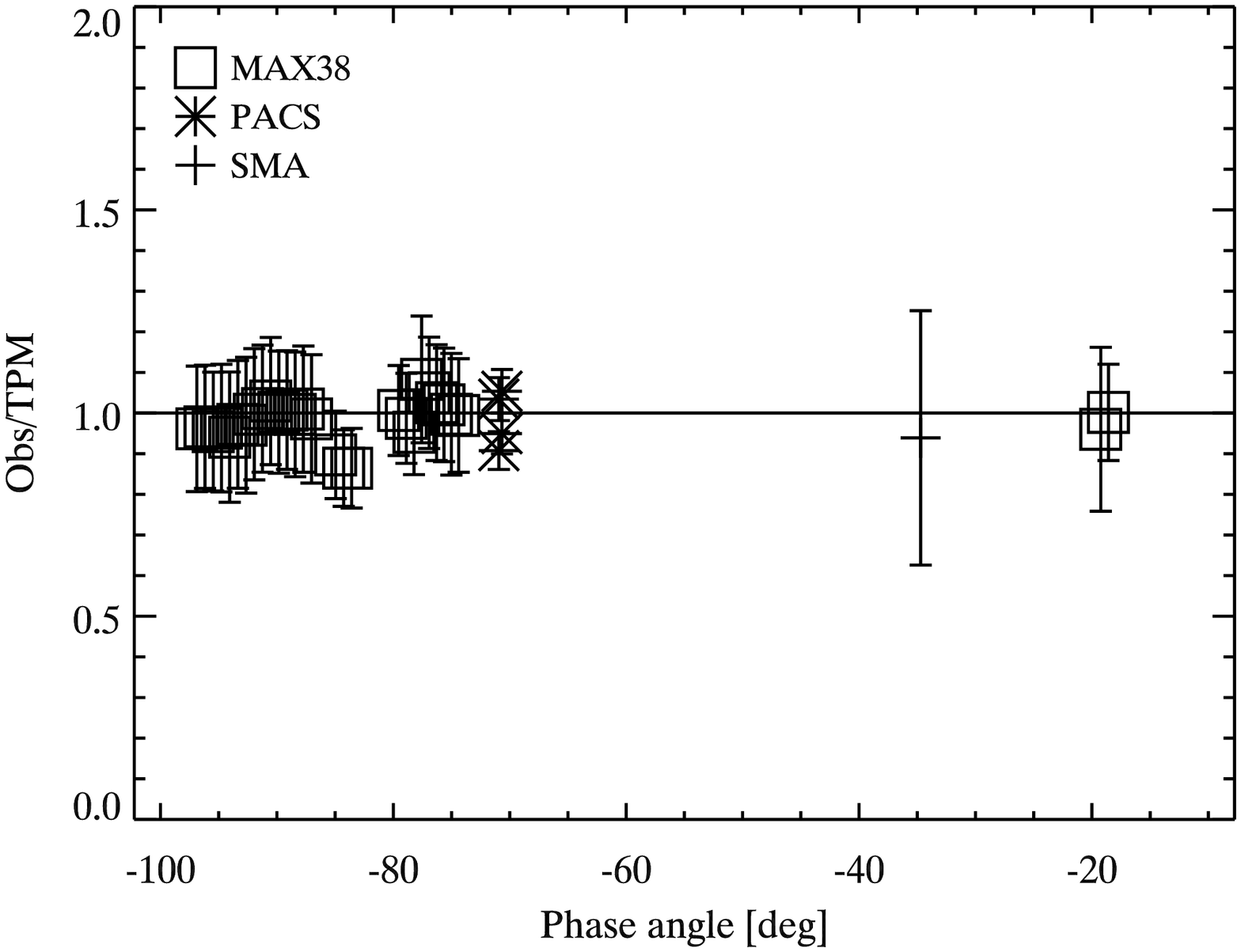}}}
 \rotatebox{90}{\resizebox{!}{10cm}{\includegraphics{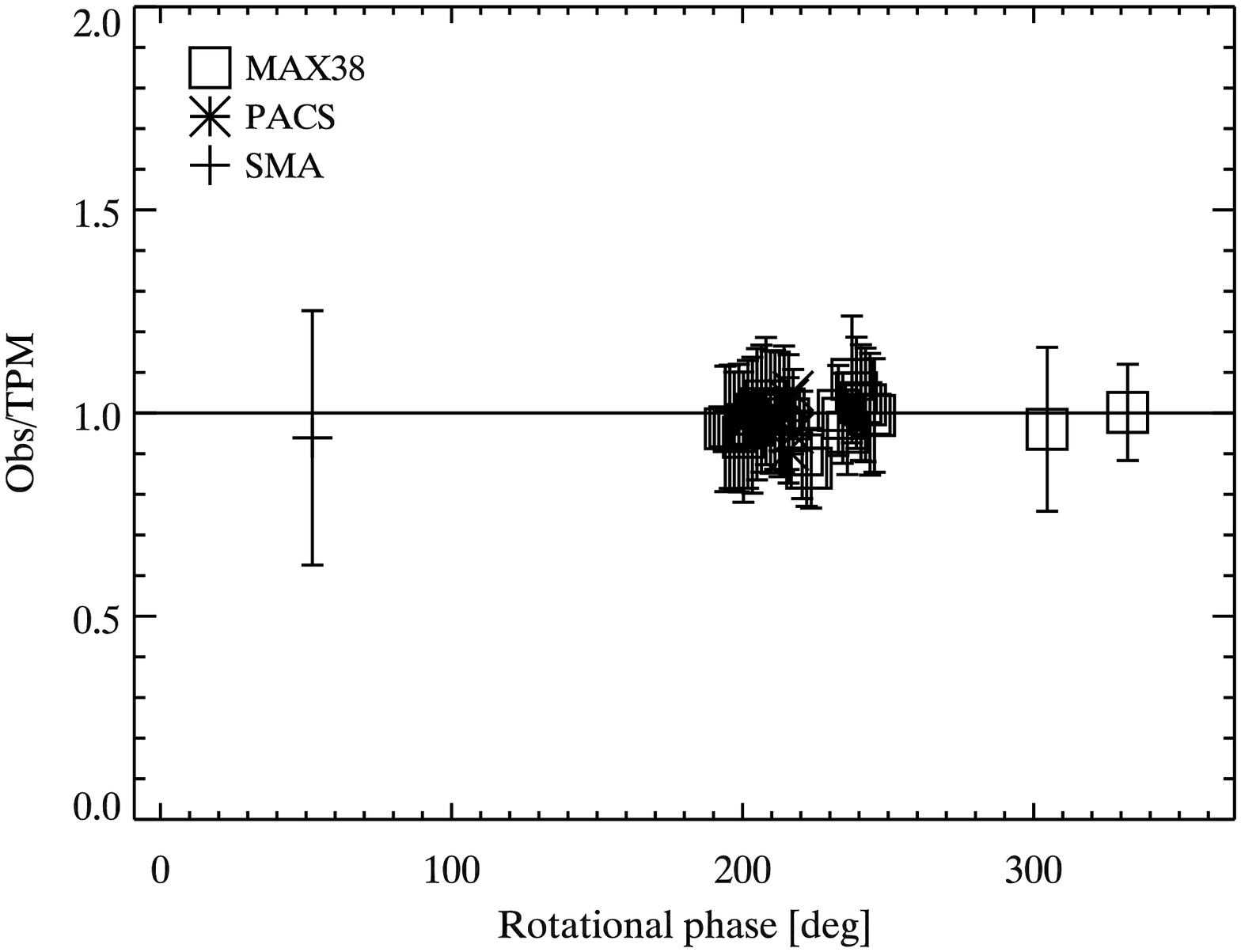}}}
  \caption{Observered and calibrated flux densities divided by the corresponding TPM prediction. Top: as a function of
           wavelength. Middle: as a function of phase angle. Bottom: as a function of rotational phase.
     \label{fig:obsmod}}
\end{figure}

Figure~\ref{fig:obsmod} also shows that 2005~YU$_{55}$ must be close 
to a sphere. An elongated or strangely shaped body would produce
a thermal lightcurve, but our dataset does not show any significant
deviations at specific rotational phases (bottom figure). But not all
rotational phases have been covered by our thermal measurements and
some of the observational errors are large.
There is also the possibility that effects of an ellipsoidal shape
could have been compensated by roughness effects (a larger cross-section
combined with a low surface roughness could produce the same flux levels
as a smaller cross-section combined with high surface roughness).
Figure \ref{fig:obsmod} (bottom) would then also show a constant
ratio at all rotational phases. But since the roughness influences the
flux in a wavelength-dependent manner (see e.g., M\"uller \cite{mueller02}, Fig.\ 3),
one should then see a larger scatter in Fig.\ \ref{fig:obsmod} (top)
at short wavelengths where the roughness has the greatest influence
on the observed fluxes. At long wavelengths (beyond $\sim$20\,$\mu$m) the
effects of roughness are much smaller and the shape effects are dominating.
Shape effects or combined shape/roughness variations are not seen in our dataset.

We did also an additional test
to see if the optical lightcurve amplitude of 0.20 $\pm$ 0.02\,mag
(Warner et al.\ \cite{warner12a}; \cite{warner12b}) is compatible with
our findings. Such an amplitude would mean that the flux at lightcurve
maximum is about 1.2 times the flux at lightcurve minimum, which would
require a SNR$>$10 time series data set for confirmation. The PACS data
are of sufficient quality, but they are taken at a single epoch. The
miniTAO/MAX38 measurements have too large error bars, related mainly to
systematic errors in the absolute flux calibration scheme. However, we
looked at the relative variation of the 22 miniTAO/MAX38 data points taken
at 18.7\,$\mu$m with respect to the spherical shape model flux predictions.
The deviations never exceed 10\%, but these data cover only a very limited
range of rotational phases (from 195 to 245$^{\circ}$ and a single point
at 305$^{\circ}$ in the bottom of figure~\ref{fig:obsmod}). The thermal
data are therefore perfectly compatible with the optical lightcurve
results and there are not indications for large deviations from a 
spherical shape.

\subsection{Error calculations}

We combine the constraints from the radar
measurements (retrograde sense of rotation, estimate of spin-axis orientation),
the AO findings (effective diameter of 307 $\pm$ 15\,m for "southern poles"), and
the speckle technique (object-averaged diameter of 285 $\pm$ 25\,m)
with the $\chi^2$ analysis for the possible spin-axis orientations
(see Figs.~\ref{fig:chi2_sv}, \ref{fig:chi2_0306} and corresponding figures
for different roughness levels which are not shown here).
For a good fit the reduced $\chi^2$-values should be close
to 1 and we estimated for our dataset that the 3-$\sigma$ confidence level for
the reduced $\chi^2$ is around 1.6. This lead to an estimated spin-axis orientation of
($\lambda_{ecl}$, $\beta_{ecl}$) = (60$^{\circ}$ $\pm$ 30$^{\circ}$, -60$^{\circ}$ $\pm$ 15$^{\circ}$).

We can use the 3-$\sigma$ threshold in reduced $\chi^2$ also for the derivation
of the corresponding size and albedo range. Figure~\ref{fig:size_albedo} shows
the size and albedo solutions for the full range of thermal inertias (from 0 to 3000\,J\,m$^{-2}$\,s$^{-0.5}$\,K$^{-1}$),
the four different levels of roughness (r.m.s.-slopes of 0.1, 0.3, 0.5, 0.8, shown with different symbols) and for all
spin-axis solutions compatible with ($\lambda_{ecl}$, $\beta_{ecl}$) = (60$^{\circ}$ $\pm$ 30$^{\circ}$, -60$^{\circ}$ $\pm$ 15$^{\circ}$).
Based on the 3-$\sigma$ confidence level we derived a possible diameter range of 295 to 322\,m,
0.057 to 0.068 for the geometric albedo, and a thermal inertia larger than 150\,J\,m$^{-2}$\,s$^{-0.5}$\,K$^{-1}$.

As a second step we looked in more details at the derived size, albedo and
thermal inertia ranges. The solutions close to the 3-$\sigma$ threshold
in our $\chi^2$-analysis are very problematic in the sense that they produce
strong trends in the observation/model figures (see Fig.~\ref{fig:obsmod}) either
with wavelengths and/or with phase angle. These kind of trends are very 
difficult to catch in an automatic $\chi^2$-analysis. We therefore moved back
to the 1-$\sigma$ solutions, corresponding to a possible diameter range of 300-312\,m,
a geometric albedo range of 0.062-0.067, and a thermal inertia range of 350-1000\,J\,m$^{-2}$\,s$^{-0.5}$\,K$^{-1}$.
The smallest thermal inertia values are connected to low roughness values (r.m.s.-slopes $\le$ 0.3)
and the largest thermal inertia values to very rough surface levels (r.m.s.-slopes $\ge$ 0.5).
The calculations for the Michelle/Gemini North data put another constraint on the 
thermal inertia and reduce the possible range to 350 - 800\,J\,m$^{-2}$\,s$^{-0.5}$\,K$^{-1}$.
The derived radiometric albedo range of 0.062-0.067 is connected
to the H$_V$ magnitude of 21.2\,m, if we include the $\pm$0.15\,mag, then the possible 
range is significantly bigger: from 0.055 to 0.075.

\begin{figure}[h!tb]
 \rotatebox{90}{\resizebox{!}{\hsize}{\includegraphics{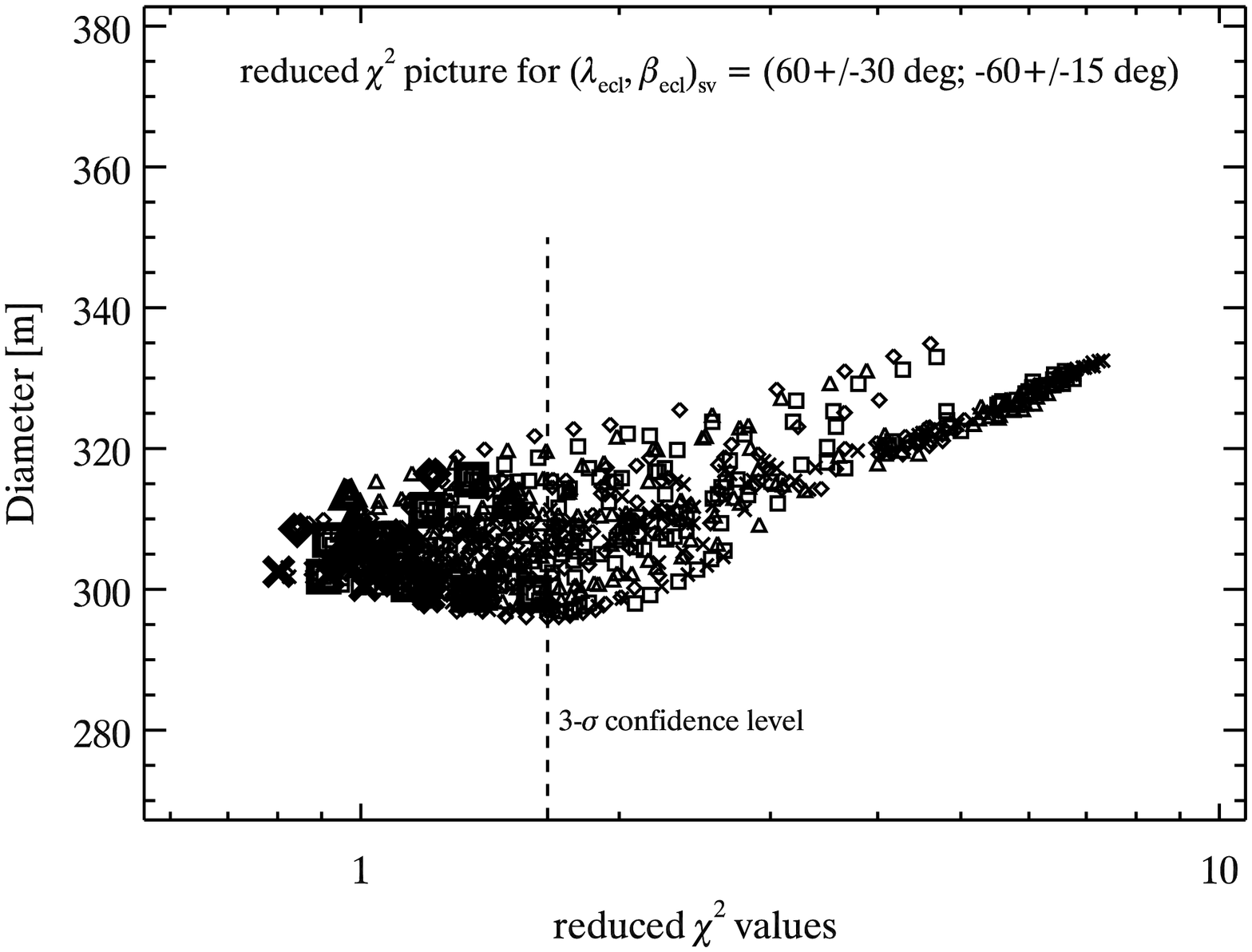}}}
 \rotatebox{90}{\resizebox{!}{\hsize}{\includegraphics{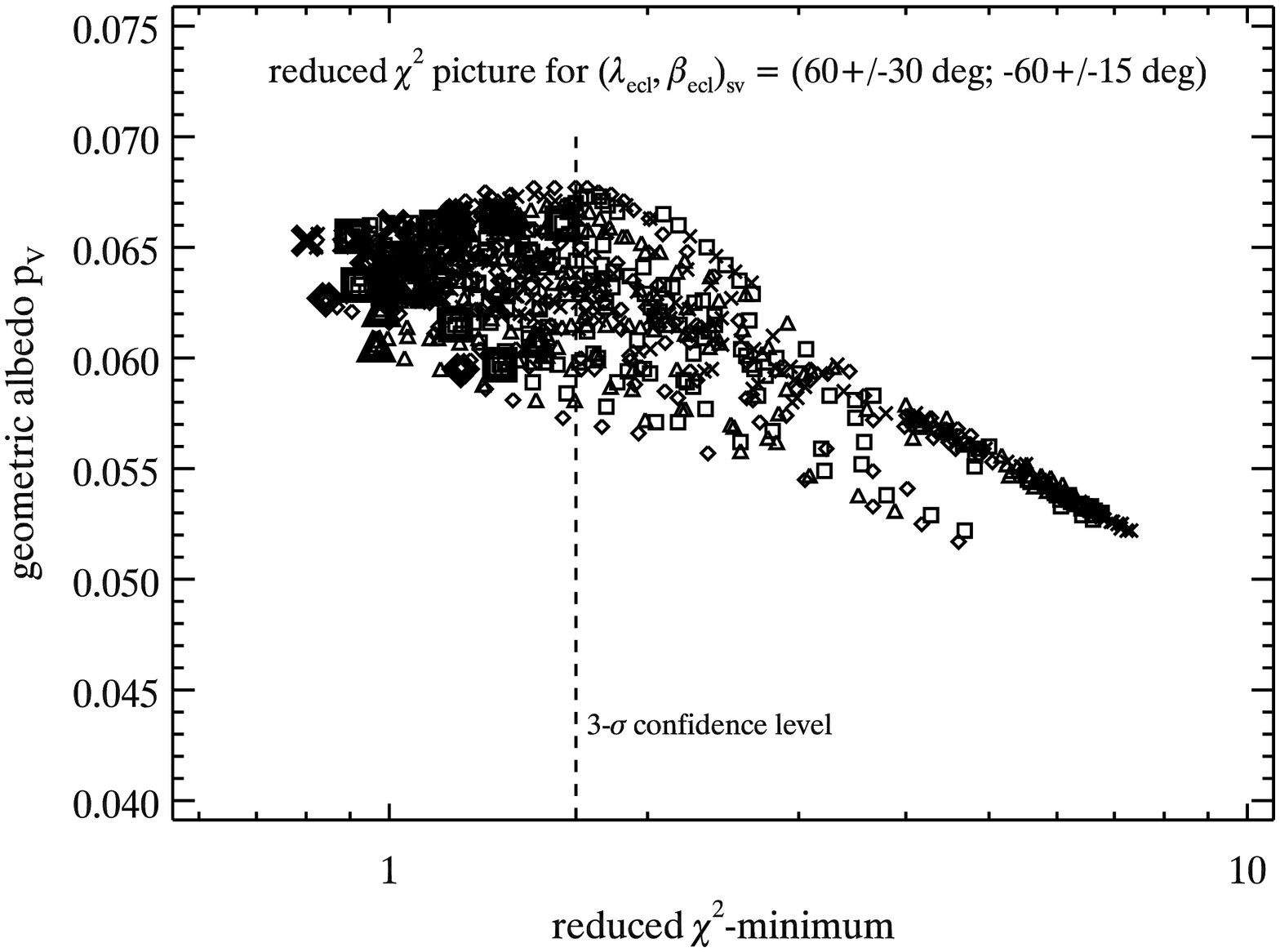}}}
  \caption{The size and albedo solutions for the full range of thermal inertias,
           the four different levels of roughness and for the most likely
	   spin-axis solutions.
     \label{fig:size_albedo}}
\end{figure}

\section{Conclusions}
\label{sec:con}

Here is a short summary of our findings for the near-Earth asteroid
2005~YU$_{55}$:
   \begin{enumerate}
      \item Our thermal data can be explained via a spherical shape model
            without seeing significant offsets at specific rotational phases,
	    showing that 2005~YU$_{55}$ is almost spherical.
      \item Our best spin-axis solution can be specified by 
           ($\lambda_{ecl}$, $\beta_{ecl}$) = (60$^{\circ}$ $\pm$ 30$^{\circ}$, -60$^{\circ}$ $\pm$ 15$^{\circ}$).
            However, the analysis of the thermal data alone would also
            allow for specific spin-axis orientations in the
            northern ecliptic hemisphere with a prograde rotation
            of the object.
      \item The radiometric analysis of our thermal data which span a wide range
            of phase angles and wavelengths (best visible in the $\chi^2$-picture 
            in Fig.~\ref{fig:chi2_0306}) is compatible with changing spin-axis
            orientations, which might be an indication for a non-principal axis rotation
            of 2005~YU$_{55}$.
      \item 2005~YU$_{55}$ has a possible effective diameter range of D$_{equ}$ = 300 - 312\,m
            (equivalent diameter of an equal volume sphere); this range was derived
	    under the assumption that the spin-axis is indeed as specified above.
      \item The analysis of all available data combined revealed a discrepancy with the radar-derived size.
      \item The geometric visual albedo p$_V$ was radiometrically derived to be in the range 0.062 to 0.067
            (H$_V$ = 21.2\,mag) or 0.055 - 0.075 if we include the $\pm$0.15\,mag error in H$_V$,
            in agreement with the C-type taxonomic classification.
      \item 2005~YU$_{55}$ has a thermal inertia in the range 350-800\,Jm$^{-2}$s$^{-0.5}$K$^{-1}$,
            very similar to the value found for the rubble-pile asteroid (25143)~Itokawa by
	    M\"uller et al.\ (\cite{mueller05}). We expect therefore that the surface of
	    2005~YU$_{55}$ looks also very similar and is composed of low conductivity fine
	    regolith mixed with larger rocks and boulders which have much higher thermal inertias.
      \item The observed thermal emission can be best reproduced when
            considering a low to intermediate roughness with an r.m.s.-slope of
            0.1-0.3; the lower roughness (or smoother surface)
            is connected to the lower thermal inertias, while a
            higher roughness would require also the higher inertia values.
   \end{enumerate}

\begin{acknowledgements}
  We would like to thank the Herschel operations team which
  supported the planning and scheduling of our fixed-time
  observations. Without their dedication and enthusiasm these
  measurements would not have been possible.
  The Submillimeter Array is a joint project between the Smithsonian
  Astrophysical Observatory and the Academia Sinica Institute of
  Astronomy and Astrophysics and is funded by the Smithsonian
  Institution and the Academia Sinica.
  SH is supported by the Space Plasma Laboratory, ISAS/JAXA.
  AP is supported by the Hungarian grant LP2012-31/2012.
\end{acknowledgements}


\begin{thebibliography}{}

\bibitem[2012]{asano12}
   Asano, K., Miyata, T., Sako, S.\ et al., 2012,
   Proc.\ of SPIE, 8446, 115

\bibitem[2011]{bodewits11}
   Bodewits, D., Campana, S., Kennea, J.\ et al.\ 2011,
   Central Bureau Electronic Telegrams, 2937, 1 (2011)

\bibitem[2012]{busch12}
   Busch, M.\ W., Benner, L.\ A.\ M., Brozovic, M.\ et al.\ 2012,
   Asteroids, Comets, Meteors 2012, conference proceedings,
   LPI Contribution No.\ 1667, id.\ 6179

\bibitem[1999]{cohen99}
   Cohen, M., Walker, R. G., Carter, B.\ et al., 1999,
   AJ, 117, 1864

\bibitem[2007]{delbo07}
   Delbo, M., dell'Oro, A., Harris, A.\ W.\ et al.\ 2007,
   Icarus 190, 236
   
\bibitem[2010]{hicks10}
  Hicks, M., Lawrence, K., Benner, L.\ 2010, The Astronomer's Telegram, 2571

\bibitem[2011]{hicks11}
   Hicks, M., Somers, J., Truong, T.\ \& Teague, S.\ 2011, The Astronomer's Telegram, 3763

\bibitem[2012]{horner12}
   Horner, J., M\"uller, T.\ G., Lykawka, P.\ S.\ 2012,
   MNRAS 423, 2587-2596

\bibitem[2012]{keihm12}
   Keihm, S., Tosi, F., Kamp, L.\ et al.\ 2012,
   Icarus 221, 395

\bibitem[1996]{lagerros96}
   Lagerros, J.\ S.\ V.\ 1996, A\&A 310, 1011

\bibitem[1997]{lagerros97}
   Lagerros, J.\ S.\ V.\ 1997, A\&A 325, 1226

\bibitem[1998]{lagerros98}
  Lagerros, J.\ S.\ V.\ 1998, A\&A 332, 1123

\bibitem[2010]{lim10}
   Lim, T.\ L., Stansberry, J., M\"uller, T.\ G.\ et al.\ 2010,
   A\&A 518, 148-152

\bibitem[2012a]{lim12a}
   Lim, L.\ F., Emery, J.\ P., Moskovitz, N.\ A., Granvik, M.\ 2012,
   43rd Lunar and Planetary Science Conference,
   LPI Contribution No.\ 1659, id.\ 2202

\bibitem[2012b]{lim12b}
   Lim, L.\ F., Emery, J.\ P., Moskovitz, N.\ A., Granvik, M.\ 2012,
   Asteroids, Comets, Meteors 2012, conference proceedings,
   LPI Contribution No.\ 1667, id.\ 6295

\bibitem[2012c]{lim12c}
   Lim, L.\ F., Emery, J.\ P., Moskovitz, N.\ A., et al.\ 2012,
   DPS meeting \#44, \#305.01

\bibitem[2011]{merline11}
   Merline, W.\ J., Drummond J.\ D., Tamblyn, P.\ M.\ et al.\ 2011,
   IAU Circular 9242, {\tt http://www.cbat.eps.harvard.edu/iauc/09200/09242.html}

\bibitem[2012]{merline12}
   Merline, W.\ J., Drummond J.\ D., Tamblyn, P.\ M.\ et al.\ 2012,
   Asteroids, Comets, Meteors 2012, conference proceedings,
   LPI Contribution No.\ 1667, id.\ 6372

\bibitem[2008]{miyata08}
   Miyata, T., Sako, S., Nakamura, T.\ et al., 2008,
   Proc.\ SPIE 7014, Ground-based and Airborne Instrumentation for Astronomy II, 701428

\bibitem[1998]{mueller98}
   M\"uller, T.\ G.\ \& Lagerros, J.\ S.\ V.\ 1998,
   A\&A, 338, 340-352

\bibitem[2002]{mueller02}
   M\"uller, T.\ G.\ 2002,
   M\&PS, 37, 1919

\bibitem[2004]{mueller04a}
   M\"uller, T.\ G.\ \& Blommaert, J.\ A.\ D.\ L.\ 2004,
   A\&A, 418, 347-356
  
\bibitem[2004]{mueller04b}
   M\"uller, T.\ G., Sterzik, M.\ F., Sch\"utz, O.\ et al.\ 2004,
   A\&A, 424, 1075-1080

\bibitem[2005]{mueller05}
  M\"uller, T.\ G., Sekiguchi, T., Kaasalainen, M.\ et al.\ 2005,
  A\&A, 443, 347-355

\bibitem[2011a]{mueller11a}
   M\"uller, T.\ G., \v{D}urech, J., Hasegawa, S.\ et al.\ 2011,
   A\&A, 525, 145

\bibitem[2011b]{mueller11b}
   M\"uller, T.\ G., Altieri, B.\ \& Kidger, M.\ 2011,
   IAU Circular 9241, {\tt http://www.cbat.eps.harvard.edu/iauc/09200/09241.html}

\bibitem[2012]{mueller12}
  M\"uller, T.\ G., O'Rourke, L., Barucci, A.\ M.\ et al.\ 2012,
  A\&A, 548, 36-45

\bibitem[2010]{nakamura10}
   Nakamura, T., Miyata, T., Sako, S.\ et al., 2010,
   Proc.\ SPIE 7735, Ground-based and Airborne Instrumentation for Astronomy III, 773561

\bibitem[2010]{nolan10}
   Nolan, M.\, C., Vervack, R.\ J., Howell, E.\ S.\ et al.\ 2010,
   American Astronomical Society, DPS meeting \#42, \#13.19,
   Bulletin of the American Astronomical Society, Vol.\ 42, p.\ 1056

\bibitem[2002]{ostro02}
   Ostro, S., Hudson, R.\ S., Benner, L.\ A.\ M.\ et al.\ 2002,
   in Asteroids III, W.\ Bottke, A.\ Cellino, P.\ Paolicchi and R.\ P.\ Binzel (eds),
   151-168

\bibitem[2005]{pravec05}
   Pravec, P., Harris, A.\ W., Scheirich, P.\ et al.\ 2005,
   Icarus 173, 108-131

\bibitem[2008]{sako08}
   Sako, S., Aoki, T., Doi, M.\ et al., 2008,
   Proc.\ SPIE 7012, Ground-based and Airborne Telescopes II, 70122T

\bibitem[2010]{somers10}
  Somers, J.M.\, Hicks, M., Lawrence, K.\ et al., 2010,
  DPS meeting \#42, \#13.16, BAAS 42, 1055

\bibitem[2012]{sridharan12}
  Sridharan, R., Girard, J. H. V., Lombardi, G., Ivanov, V. D., Dumas, C.\ 2012,
  Optical and Infrared Interferometry III. Proceedings of the SPIE, Volume 8445
  
\bibitem[2012a]{taylor12a}
   Taylor, P.\ A., Nolan, M.\ C., Howell, E.\ S.\ et al.\ 2012,
   American Astronomical Society, AAS Meeting \#219, \#432.11

\bibitem[2012b]{taylor12b}
   Taylor, P.\ A., Howell, E.\ S., Nolan, M.\ C.\ et al.\ 2012,
   Asteroids, Comets, Meteors 2012, conference proceedings,
   LPI Contribution No.\ 1667, id.\ 6340

\bibitem[2010]{vodniza10}
  Vodniza, A.Q. \& M.R. Pereira, 2010,
  DPS meeting \#42, \#13.25, BAAS 42, 1057

\bibitem[2012a]{warner12a}
   Warner, B.\ D., Stephens, R.\ D., Brinsfield, J.\ W.\ et al.\ 2012,
   The Minor Planet Bulletin, Association of Lunar and Planetary Observers,
   Vol.\ 39, No.\ 2, 84-85

\bibitem[2012b]{warner12b}
   Warner, B.\ D., Stephens, R.\ D., Brinsfield, J.\ W.\ et al.\ 2012,
   Asteroids, Comets, Meteors 2012, conference proceedings,
   LPI Contribution No.\ 1667, id.\ 6013

\bibitem[2010]{yoshii10}
   Yoshii, Y., Aoki, T., Doi, M.\ et al.\ 2010,
   Proc.\ SPIE 7733, Ground-based and Airborne Instrumentation for Astronomy III, 773308


\end{thebibliography}
\end{document}